\documentclass[11pt,a4paper]{article}
\pdfoutput=1
\usepackage{jcappub}

\usepackage{bm,graphicx,amssymb}

\begin{document}

\title{Power spectrum tomography of dark matter annihilation with 
local galaxy distribution}
\author{Shin'ichiro Ando}

\affiliation{GRAPPA Institute, University of Amsterdam, 1098 XH
Amsterdam, The Netherlands}

\emailAdd{s.ando@uva.nl}

\abstract{Cross-correlating the gamma-ray background with local galaxy
catalogs potentially gives stringent constraints on dark matter
annihilation.
We provide updated theoretical estimates of sensitivities to the
annihilation cross section from
gamma-ray data with Fermi telescope and 2MASS galaxy catalogs, by
elaborating the galaxy power spectrum and astrophysical backgrounds, and
adopting the Markov-Chain Monte Carlo simulations.
In particular, we show that taking tomographic approach by
dividing the galaxy catalogs into more than one redshift slice will
improve the sensitivity by a factor of a few to several.
If dark matter halos contain lots of bright substructures, yielding a
large annihilation boost (e.g., a factor of $\sim$100 for galaxy-size halos),
then one may be able to probe the canonical annihilation cross section
for thermal production mechanism up to masses of $\sim$700~GeV.
Even with modest substructure boost (e.g., a factor of $\sim$10 for
galaxy-size halos), on the other hand, the
sensitivities could still reach a factor of three larger than the
canonical cross section for dark matter masses of tens to a few hundreds
of GeV.}
\maketitle

\section{Introduction}
\label{sec:introduction}

If dark matter is made of weakly interacting massive particles, as
suggested by popular class of particle physics
models~\cite{Jungman1996}, they may self-annihilate and leave
observable signatures in the high-energy sky.
Since the annihilation rate scales as density squared, searches for
high-energy radiations such as gamma rays have been performed towards
various dense regions of the sky, having yielded no definite signatures
yet~\cite{Bringmann2012}.

Since the gamma rays from dark matter annihilation precisely trace the
density squared of dark matter particles, it has been proposed to study
the {\it anisotropy} in the diffuse gamma-ray background to search for
a characteristic signature of the dark matter annihilation~\cite{AK2006,
Ando2007, Miniati2007, Cuoco2008, SiegalGaskins2008, Lee2008,
Fornasa2009, Ando2009, Zavala2009, Ando2009b, Ibarra2010,
SiegalGaskins2011, Cuoco2011, Fornasa2013, AK2013}. 
Reference~\cite{FermiAnisotropy} analyzed the 22-month gamma-ray data
from the Fermi Large Area Telescope (LAT), and detected excess in
angular power spectrum over the shot noise of the photons.
This excess is then interpreted to be consistent with what one expects
from unresolved blazar contributions~\cite{FermiAnisotropy,
Cuoco2012}.
No signature of dark matter annihilation has been found yet,
although interesting upper limits on annihilation cross section were
obtained~\cite{AK2013, Gomez-Vargas2013}.

As the dark matter distribution is well traced by galaxies,
Ref.~\cite{ABK} showed that by {\it cross-correlating} the gamma-ray map
with local galaxy catalogs, one could vastly improve sensitivities on
the dark matter annihilation.
(See also Ref.~\cite{Fornengo2014}.)
In particular, nearby galaxy catalogs such as Two Micron All-Sky Survey
(2MASS)~\cite{Huchra2012} are the most ideal ones, because the most of
the contributions from dark matter annihilation come from the relatively
nearby Universe.
Furthermore, contributions from other astrophysical sources such as
star-forming galaxies (SFGs) and blazars will be suppressed, as they are
more important at higher redshifts, and therefore, are less strongly
correlated with the 2MASS catalogs.
The expected upper limits on the annihilation cross section depend on
the substructure boost just as any other extragalactic constraints, but
for an optimistic boost model~\cite{Gao2012}, they are sensitive to the
`canonical' cross section, $\langle \sigma v \rangle = (2$--$3) \times
10^{-26}$~cm$^{3}$~s$^{-1}$~\cite{Jungman1996, Steigman2012}, for dark
matter masses smaller than a few hundreds of GeV.
For a more conservative boost scenario~\cite{Sanchez-Conde2014}, the
sensitivity will be weakened by about an order of magnitude.
It has also been proposed to cross correlate the gamma-ray background
and weak gravitational lensing, in order to further improve
sensitivities for future lensing surveys~\cite{Camera2013}.

In this paper, we aim at making further theoretical developments of the
cross correlation with the galaxy catalogs.
In particular, we fully utilize information of redshifts of galaxies in
the catalog as well as energies of photons.
Specifically, we improve on the following points compared with our
previous study~\cite{ABK}.
\begin{enumerate}
\item {\it Tomographic study of the cross correlation.---}%
Galaxies in existing catalogs are in most cases assigned with
      either spectroscopic or photometric redshifts.
Since the different redshift ranges contribute differently to the
      gamma-ray background depending on the sources, one could use
      redshift information of the catalog galaxies to further
      disentangle dark matter signals from the astrophysical
      backgrounds.
Here, we divide the catalogs into several redshift bins to study
      sensitivities of such an approach.
We show as the result that one can improve the sensitivity further by a
      factor of a few to several if we divide the sample into more than
      a few redshift bins.
The same idea was mentioned for cross correlation with future lensing
      data in Ref.~\cite{Fornengo2014}.

\item {\it Improved sensitivity study.---}%
First, we include the shot noise in the cross-power spectrum, which
      comes from the fact that some catalog galaxies contribute to the
      gamma-ray background as discrete point sources.
Second, we adopt the Bayesian statistics (e.g., \cite{Trotta2008})
      and Markov-Chain Monte Carlo simulations
      (MCMC), rather than relying on a simplistic Fisher matrix approach.
This enables to use the prior information for theoretical parameters,
      and one can avoid any unphysical values of them such as negative
      values for annihilation cross section.
This effect was not well treated in the previous study, and indeed, we
      find that the expected sensitivity is improved compared to the 
      one from simple Fisher estimates.
Third, we show that the expected sensitivity has an uncertainty range of
      about one order of magnitude, coming from intrinsic statistical
      fluctuations of data.

 \item {\it Better modeling on galaxy power spectrum and
       cross-correlation power spectrum.---}%
We adopt the halo occupation distribution (HOD) within the context of
       `halo model'~\cite{Seljak2000, Cooray2002}.
The galaxy power spectrum and blazar-galaxy cross-power spectrum
       computed this way differ substantially at small scales
       from the matter power spectrum.
The cross power spectrum between density squared (for the dark matter
       annihilation) and galaxy distribution is computed in a similar
       way, and again, shows deviation at small scales from the
       cross power between density squared and density.
\end{enumerate}

This paper is organized as follows.
In Sec.~\ref{sec:Diffuse gamma-ray background}, we briefly summarize
models of both dark matter distribution and astrophysical sources, and
give formulation and estimates of the diffuse gamma-ray background.
In Sec.~\ref{sec:cross correlation}, we discuss formulation and results
of the cross-correlation power spectrum, and in
Sec.~\ref{sec:sensitivity}, we perform estimates of the sensitivity to
the annihilation cross section, based on MCMC.
Finally, we conclude the paper by giving summaries in
Sec.~\ref{sec:conclusions}.
Throughout the paper, we assume the cosmological model with cold dark
matter and cosmological constant ($\Lambda$CDM), and cosmological
parameters adopted are $\Omega_{\rm m} = 0.27$, $\Omega_{\Lambda} =
0.73$, $H_0$ = 100~$h$~km~s$^{-1}$~Mpc$^{-1}$ with $h = 0.7$, $n_s =
0.96$, and $\sigma_8 = 0.8$.

\section{Diffuse gamma-ray background}
\label{sec:Diffuse gamma-ray background}

This section introduces several quantities that are important for
computations of the gamma-ray background intensity.
We summarize it rather briefly here, but we refer the reader to our
previous papers~\cite{Ando2007, AK2013} for a more detailed discussion.

\subsection{Dark matter annihilation}

The intensity of the diffuse gamma-ray background due to dark matter
annihilation is proportional to the line-of-sight integral of dark
matter density squared along a direction $\hat{\bm n}$.
We write this as
\begin{equation}
 I_{\rm dm}(\hat{\bm n}) = \int d\chi W_{\rm dm}(z) 
  \left[\frac{\rho_{\rm dm}(\chi \hat{\bm n}, z)}{\langle\rho_{\rm
   dm} \rangle}\right]^2
  = \int d\chi W_{\rm dm}(z) [1+\delta(\chi\hat{\bm n}, z)]^2,
  \label{eq:I_dm}
\end{equation}
where the subscript `dm' represents dark matter, $\chi$ is the comoving
distance, $z$ is the redshift corresponding to $\chi$, $\rho_{\rm
dm}(\chi \hat{\bm n})$ is the (comoving) dark matter density, and
$\langle\rho_{\rm dm} \rangle = \Omega_{\rm dm} \rho_{\rm c}$ with dark
matter density parameter, $\Omega_{\rm dm} = 0.23$, and the critical
density at the present Universe, $\rho_{\rm c}$.
In the second equality, we introduced the (dark) matter overdensity
$\delta = (\rho_{\rm m} - \langle \rho_{\rm m} \rangle) / \langle
\rho_{\rm m}\rangle = (\rho_{\rm dm} - \langle \rho_{\rm dm} \rangle) /
\langle \rho_{\rm dm}\rangle$.
All the particle physics parameters such as dark matter mass $m_{\rm
dm}$, annihilation cross section $\langle \sigma v \rangle$, and
gamma-ray spectrum per annihilation $dN_{\gamma, {\rm ann}} / dE$ are
included in the window function $W_{\rm dm}$:
\begin{equation}
 W_{\rm dm}(z) = \frac{\langle \sigma v \rangle}{8\pi}
  \left(\frac{\Omega_{\rm dm}\rho_c}{m_{\rm dm}}\right)^2(1+z)^3
  \int_{E_{\rm min}}^{E_{\rm max}} dE
  \left.\frac{dN_{\gamma, {\rm ann}}}{dE^\prime}\right|_{E^\prime =
  (1+z) E} e^{-\tau(E,z)},
  \label{eq:W_dm}
\end{equation}
where $E_{\rm min}$ and $E_{\rm max}$ are the minimum and maximum
energies of the photons considered, and $\tau (E, z)$ takes into account
the absorption of the gamma rays due to interactions with the
extragalactic background light (e.g., \cite{Gilmore2012}).

We obtain the mean intensity by taking the ensemble average of
Eq.~(\ref{eq:I_dm}):
\begin{equation}
 \langle I_{\rm dm}\rangle = \int d\chi W_{\rm dm}(z)
  [1 + \langle \delta^2(\chi)\rangle].
  \label{eq:mean intensity}
\end{equation}
The variance of overdensity $\langle \delta^2(z) \rangle$ is computed as
(e.g., \cite{AK2006})
\begin{eqnarray}
 1 + \langle \delta^2(z)\rangle &=&
  \int dM \frac{dn (M,z)}{dM} \frac{\mathcal J(M,z)}{(\Omega_{\rm
  m}\rho_{\rm c})^2},
  \label{eq:variance}
 \\
 \mathcal J(M,z) &\equiv& [1 + b_{\rm sh}(M, z)]
  \int dV \rho^2_{\rm host} (r | M),
\end{eqnarray}
where $dn / dM$ is the halo mass function, $\rho_{\rm host}(r | M)$ is
the density profile of the host halo (i.e., smooth component), and
$b_{\rm sh}(M, z)$ is the `boost factor' due to presence of halo
substructures.
Note that the density $\rho_{\rm host}$ as well as the radial coordinate
$r$ are all comoving quantities.
This calculation is based on the fact that the dark matter particles are
confined in halos and their subhalos, as numerical simulations
show~\cite{Klypin1999, Moore1999}, and hence the density squared is
boosted due to enhanced clumpiness.
We adopt the Navarro-Frenk-White (NFW) density profile~\cite{NFW} for
the host halos, and with the fit, the volume integral of the density
squared can be carried out analytically (see, e.g.,
Ref.~\cite{AK2013}).
For the halo mass function, we adopt the ellipsoidal collapse model by
Ref.~\cite{ST} down to a minimum mass of 10$^{-6} M_\odot$.
The subhalo boost factor $b_{\rm sh}(M)$ yields the largest
uncertainties in this computation.
In order to bracket the uncertainty, we adopt two models: one model that
extrapolates the power-law dependence of the boost on subhalo masses to
the smallest halo mass by Ref.~\cite{Gao2012}, and the other relying on
physically motivated models that well reproduce concentration parameters
of the field halos but applied to the subhalos~\cite{Sanchez-Conde2014}.

\begin{figure}[t]
 \begin{center}
  \includegraphics[height=6cm]{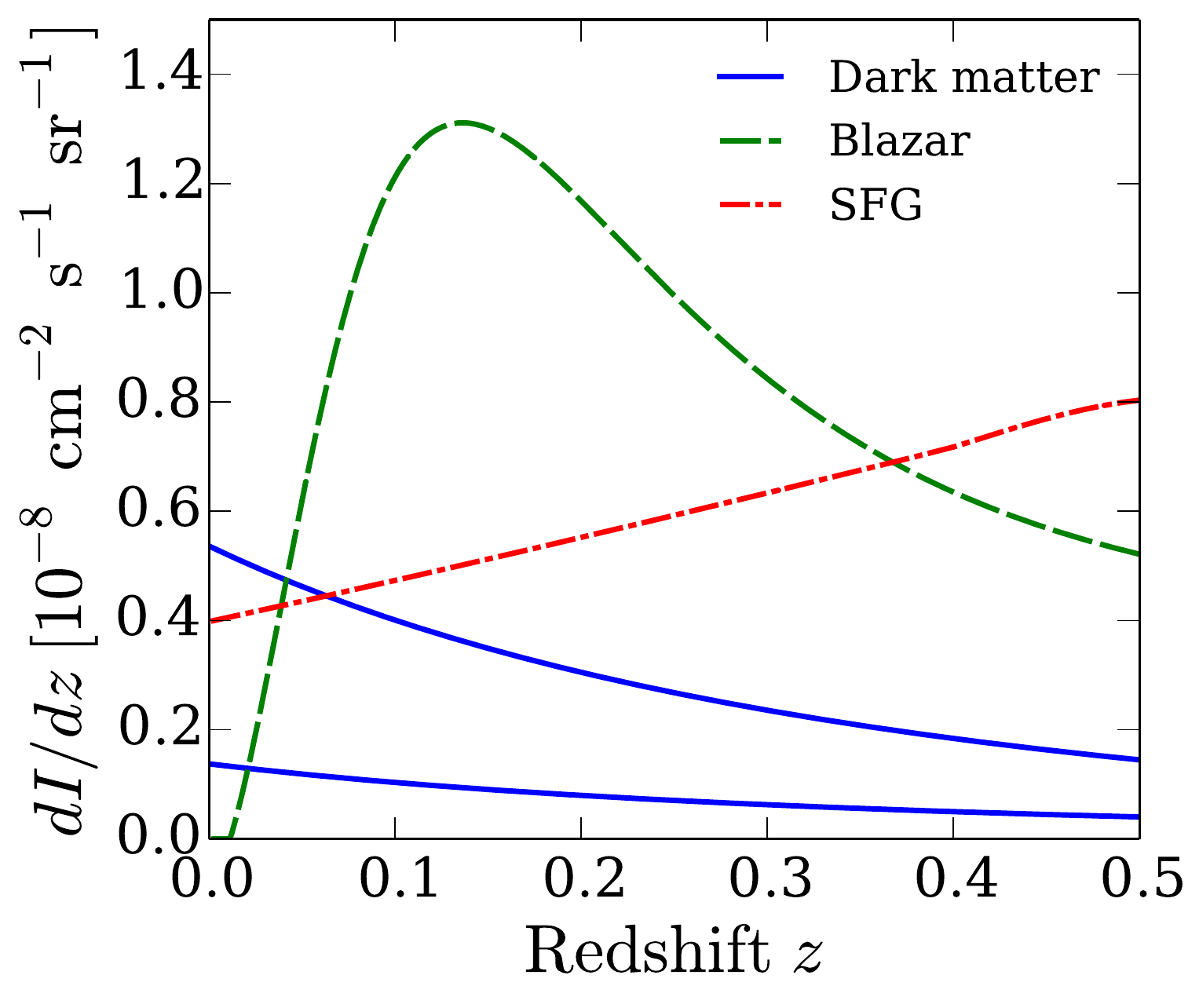}
  \includegraphics[height=6cm]{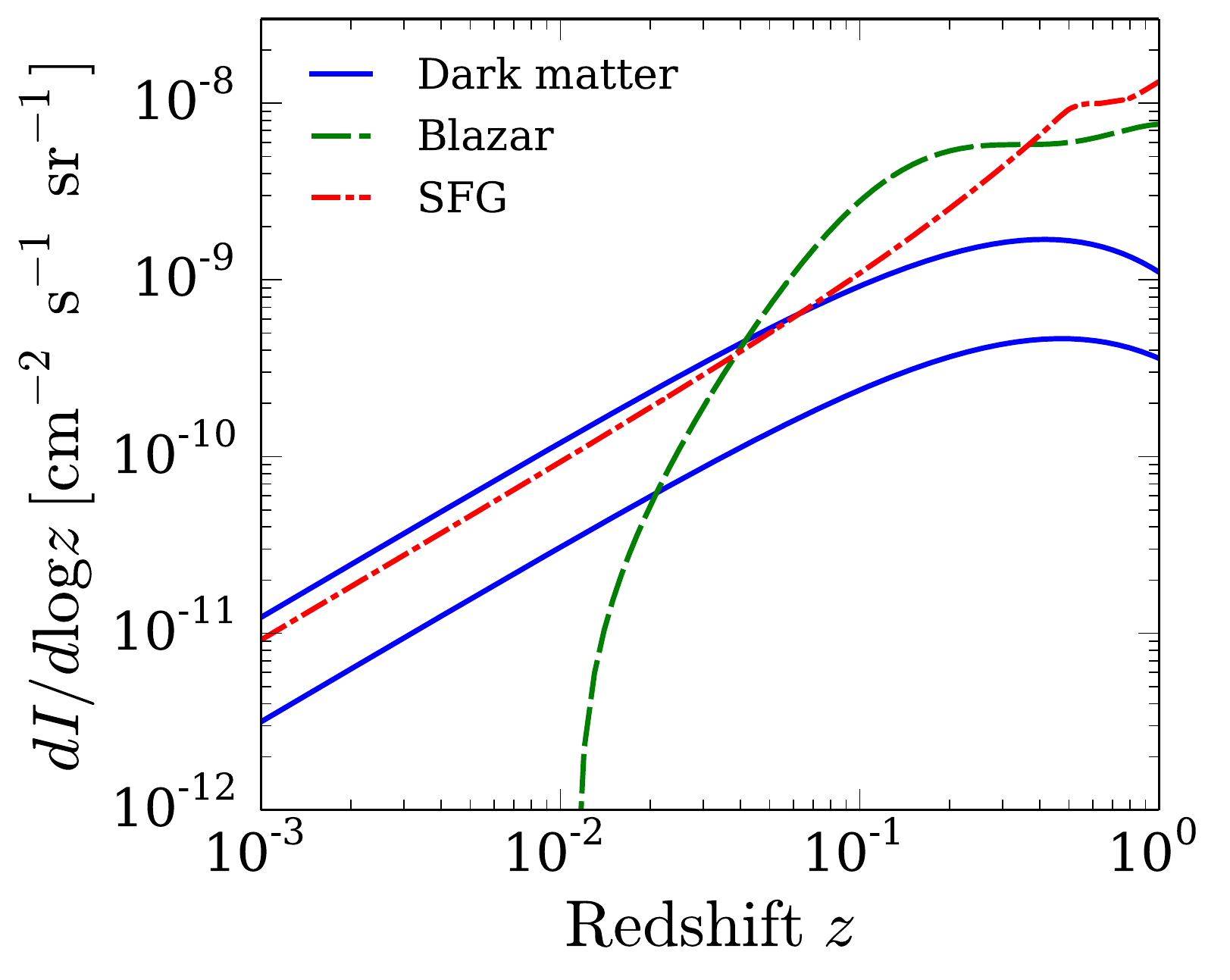}
  \caption{Contribution to the diffuse gamma-ray intensity from
  different redshift ranges, $dI / dz$ (left) and $dI / d\log z$
  (right), for the energy band of
  5--10~GeV. For dark matter, $m_{\rm dm} = 100$~GeV, $\langle \sigma v
  \rangle = 3 \times 10^{-26}$~cm$^3$~s$^{-1}$, and $b\bar b$
  annihilation channel are assumed. The upper and lower solid curves
  correspond to the boost models~\cite{Gao2012, Sanchez-Conde2014},
  respectively.}
  \label{fig:dIdz}
 \end{center}
\end{figure}

\begin{figure}[t]
 \begin{center}
  \includegraphics[width=8.5cm]{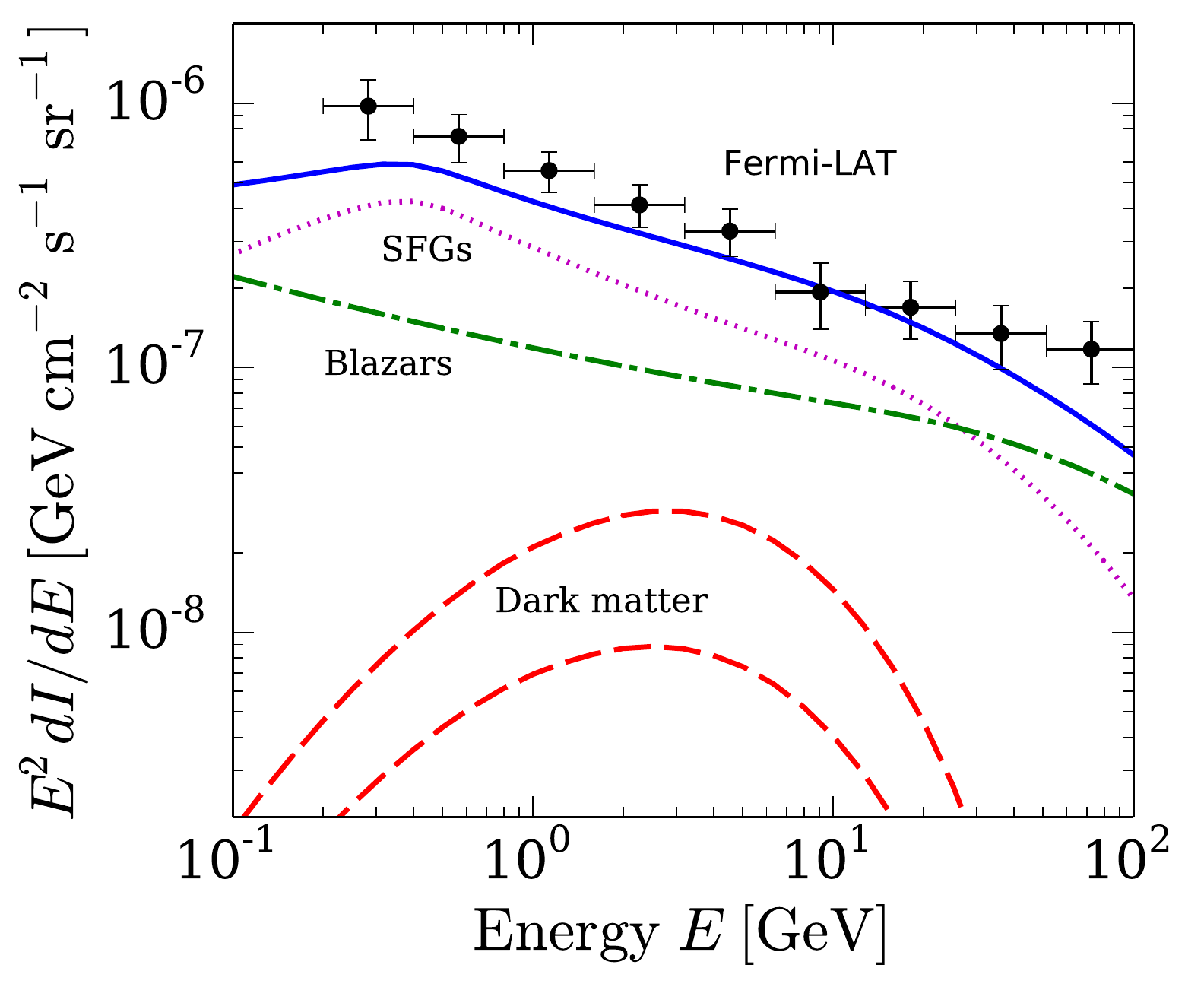}
  \caption{Spectrum of the diffuse gamma-ray background $E^2 dI / dE$,
  compared with the Fermi-LAT data~\cite{FermiDiffuse}. The models of
  dark matter, SFGs, and blazars are the same as in
  Fig.~\ref{fig:dIdz}, and the total contribution is shown as a solid
  curve.}
  \label{fig:spectrum}
 \end{center}
\end{figure}

Figure~\ref{fig:dIdz} shows contributions to the mean intensity from
different redshift ranges, $dI_{\rm dm} / dz \propto W_{\rm dm}(z)
[1 + \langle \delta^2(z)\rangle]$, for energy band of 5--10~GeV, and the
two boost models~\cite{Gao2012, Sanchez-Conde2014}.
In this figure and also in the followings (unless stating otherwise), we
assume the canonical dark matter parameters: $m_{\rm dm} = 100$~GeV,
$\langle \sigma v\rangle = 3\times 10^{-26}$~cm$^3$~s$^{-1}$, and
annihilation channel purely into $b\bar b$.
The dark matter contribution mainly comes from the low-redshift regime,
in particular for $z \lesssim 1$.
In Fig.~\ref{fig:spectrum}, we show the spectrum of the mean intensity
after integrating over all redshifts.
The shape of the energy spectrum of the dark matter component is
characteristic featuring bump at fraction of dark matter mass, although
the amplitude is smaller than the astrophysical components or the
Fermi-LAT data~\cite{FermiDiffuse}, for the dark matter parameters
adopted here.

\subsection{Astrophysical sources}
\label{sub:Astrophysical sources}

The gamma-ray intensity due to an astrophysical source population $X$ is
proportional to the line-of-sight integral of its number density $n_X$
as
\begin{equation}
 I_X (\hat{\bm n}) = \int d\chi W_X(z)\frac{n_X(\chi\hat{\bm n},
  z)}{\langle n_X(z)\rangle} = \int d\chi W_X(z) [1 +
  \delta_X(\chi\hat{\bm n}, z)],
  \label{eq:I_X}
\end{equation}
where, as in Eq.~(\ref{eq:I_dm}), we introduced the overdensity of the
source $X$ as $\delta_X = (n_X - \langle n_X \rangle) / \langle n_X
\rangle$.
Since $\langle \delta_X \rangle = 0$, the mean intensity $\langle I_X
\rangle$ is simply obtained by the integration of $W_X$ over $\chi$.

Often in the literature, the luminosity function is constructed from
observational data to represent the (comoving) number density of the
source per unit luminosity range.
The luminosity is defined as the number of gamma-ray photons emitted per
unit time in the source rest frame.
The window function $W_X$ is written in terms of the luminosity function
$\Phi_X(\mathcal L, z)$ as
\begin{equation}
 W_X(z) = \chi^2 \int_{E_{\rm min}}^{E_{\rm max}}dE
  \int_0^{\mathcal L_{\rm lim}} d\mathcal L \Phi_X (\mathcal L, z)
  \mathcal F(\mathcal L, z),
\end{equation}
where $\mathcal L$ is the differential number luminosity (number of the
photons emitted per unit time per unit energy range) at energy of $(1+z)
E$ and $\mathcal F = \mathcal L / (4 \pi \chi^2)$ is the differential
photon flux received at energy $E$.
The upper limit of the luminosity integral $\mathcal L_{\rm lim}$
corresponds to the sensitivity flux to point sources $F_{\rm sen}$,
above which the sources are recognized as resolved and hence do not
contribute to the diffuse background.
These are related to each other through
\begin{equation}
 F_{\rm sen} = \frac{1}{4\pi \chi^2}\int dE \mathcal L_{\rm
  lim}([1+z]E, z),
\end{equation}
and it is straightforward to invert this relation for $\mathcal L_{\rm
lim}$ once the spectral index is specified (see below).

As astrophysics sources, in this paper, we consider blazars and SFGs;
i.e., $X = \{{\rm b}, {\rm SFG}\}$, where `b' stands for the blazars.
These two are representative source classes that are discussed in the
literature most extensively, and also their gamma-ray luminosity
functions are relatively well established (in particular for the
blazars).
Other extragalactic sources such as radio galaxies or mis-aligned active
galactic nuclei (AGNs) might also give substantial contributions, although
uncertainties are much larger~\cite{DiMauro2014}.
The main conclusion from the cross-correlation analyses reached in the
subsequent sections, however, will be largely unaffected by removing
these sources from our consideration.
This is mainly because our approach in Sec.~\ref{sec:sensitivity} is
model independent in a sense that the amplitude of the cross-power
spectrum of the astrophysical components are set as free parameters.
Therefore, as also discussed in Ref.~\cite{ABK}, unless these other
sources such as mis-aligned AGNs feature an energy spectrum or clustering
property very different from those of conventional astrophysical sources
(that is unlikely), these components will be degenerate with either
blazars or SFGs.

The blazars are further divided into two sub-classes: flat-spectrum
radio quasars (FSRQs) and BL Lac objects.
The gamma-ray luminosity functions of the both populations are well
established from the direct measurements, and we adopt the
luminosity-dependent density evolution models~\cite{Ajello2012,
Ajello2014}.
The blazar luminosity functions are well represented with a double
power law, and we impose the lower luminosity cutoff at
10$^{42}$~erg~s$^{-1}$.
The spectrum of FSRQs is well approximated by a power law with an
index of $2.44$, whereas that of BL Lacs is by that with an index of
$2.1$~\cite{Abdo2010a, Abdo2010b}.
For these soft and hard sources, we adopt the sensitivity flux
(integrated above 100~MeV) of $F_{\rm sen} = 3 \times
10^{-8}$~cm$^{-2}$~s$^{-1}$ and $4 \times 10^{-9}$~cm$^{-2}$~s$^{-1}$,
respectively~\cite{Abdo2010b}.

For the SFGs, we adopt the luminosity function of galaxies measured in
the infrared waveband with Herschel~\cite{Gruppioni2013}, following
Ref.~\cite{Tamborra2014}.
We treat the contributions due to normal spiral galaxies and
starburst galaxies separately, where the gamma-ray spectrum is
approximated as $E^{-2.7}$ and $E^{-2.2}$ for the former and latter
populations, respectively.
The sensitivity flux for these populations are again $F_{\rm sen} = 3 \times
10^{-8}$~cm$^{-2}$~s$^{-1}$ and $4 \times 10^{-9}$~cm$^{-2}$~s$^{-1}$,
respectively.
The infrared luminosity function is then converted to the gamma-ray
luminosity function through the correlation between infrared luminosity
and gamma-ray luminosity of the galaxies~\cite{Ackermann2012}:
\begin{equation}
 \log\left(\frac{L_\gamma}{\mathrm{erg~s^{-1}}}\right) = 1.17 \log
  \left(\frac{L_{\rm IR}}{10^{10} L_\odot}\right) + 39.28,
\end{equation}
where $L_\gamma$ is the gamma-ray luminosity integrated for 0.1--100~GeV
and $L_{\rm IR}$ is the infrared luminosity for 8--1000~$\mu$m.

Figure~\ref{fig:dIdz} also shows the astrophysical components for $dI /
dz$.
One can see that, unlike dark matter annihilation, these components
increase with the redshifts following the trend in their luminosity
functions.
In particular, since blazars are bright individually (they are assumed
to be brighter than 10$^{42}$~erg~s$^{-1}$ above 100~MeV), all of them
within $z = 0.01$ will be resolved, and there is no contribution from
this component.
SFGs are, on the other hand, harder to resolve individually as they are
much less luminous, although they contribute to the gamma-ray background
at a comparable level to the blazars.
The diffuse gamma-ray background takes into account all the redshift
contributions, and therefore, it is very difficult for dark matter
component to excel unless its annihilation cross section is much
larger than $3 \times 10^{-26}$~cm$^{3}$~s$^{-1}$
(Fig.~\ref{fig:spectrum}).
However, if one can single out nearby redshift regime, the dark matter
component can be the dominant one, and this is realized by taking cross
correlation with the local galaxy catalogs as discussed in
Ref.~\cite{ABK} and also in the following section.

\section{Cross correlation between gamma-ray background and galaxy
 catalogs}
\label{sec:cross correlation}

\subsection{Galaxy catalogs}

As we showed in Sec.~\ref{sub:Astrophysical sources}, we are interested
in extracting low-redshift information in the gamma-ray background.
In the local Universe, 2MASS catalog provides nearly complete
information on the galaxy distribution.
In particular, 2MASS Redshift Survey (2MRS) is based on spectroscopic
redshift determination of $\sim$43500 galaxies up to $z \sim 0.1$ from
almost all the sky (sky coverage is $f_{\rm 2MRS} \simeq
0.91$)~\cite{Huchra2012}. 
We approximate its redshift distribution as
\begin{equation}
 \frac{dN_{\rm 2MRS}}{dz} \propto z \exp
  \left[- \left(\frac{z}{0.033}\right)^2\right],
\end{equation}
where the constant of proportionality is
computed such that it gives $N_{\rm 2MRS} = 43500$ after integration
over redshifts.\footnote{This is a simple phenomenological model that
only roughly reproduces the redshift distribution of 2MRS galaxies.
For the actual data analysis and its interpretation, one should instead
adopt a more accurate model~\cite{Branchini2012}.}
We then define the 2MRS galaxy window function as $W_{\rm 2MRS}(z) =
N_{\rm 2MRS}^{-1}(dN_{\rm 2MRS}/dz)(dz / d\chi)$, which gives unity
after integration over the comoving distance $\chi$.

We also adopt the 2MASS Extended Source Catalog (2MXSC) as even larger
sample that contains 2MRS, but with less accurate redshift determination
through photometry~\cite{Jarrett2000, Jarrett2004}.
This catalog was used for the first analysis of the cross correlation of
the gamma-ray background in Ref.~\cite{Xia2011}.
The catalog contains $N_{\rm 2MXSC} \simeq 770000$ galaxies from the
$4\pi f_{\rm 2MXSC}$-sr sky region, where $f_{\rm 2MXSC} =
0.67$~\cite{Jarrett2000, Xia2011}.
The redshift distribution of the catalog galaxies is
\begin{equation}
 \frac{dN_{\rm 2MXSC}}{dz} \propto z^{1.9} \exp
  \left[-\left(\frac{z}{0.07}\right)^{1.75}\right],
\end{equation}
and we define $W_{\rm 2MXSC}(z)$ similarly to that for the 2MRS
catalog.
We note that using extraction criteria less conservative than that of
Ref.~\cite{Xia2011} will increase $f_{\rm 2MXSC}$ significantly, and
hence the eventual sensitivity to dark matter.

\subsection{Angular cross-power spectrum: Dark matter annihilation}

The angular cross-power spectrum between the gamma-ray intensity due to
dark matter annihilation $I_{\rm dm}(\hat{\bm n})$ and the galaxy
surface density $\Sigma_{\rm g}(\hat{\bm n})$ (where this is normalized
to $\langle \Sigma_{\rm g} \rangle = 1$) is obtained as
\begin{equation}
 C_{\rm dm, g}(\theta) \equiv
 \langle \delta I_{\rm dm}(\hat{\bm n}) \delta \Sigma_{\rm g}(\hat{\bm
  n} + {\bm \theta}) \rangle = \sum_\ell \frac{2\ell + 1}{4\pi}
  C_\ell^{\rm dm, g} P_\ell (\cos \theta),
\end{equation}
where $\delta I_{\rm dm}(\hat{\bm n}) = I_{\rm dm}(\hat{\bm n}) -
\langle I_{\rm dm} \rangle$, $\delta \Sigma_{\rm g}(\hat{\bm n}) =
\Sigma_{\rm g}(\hat{\bm n}) - \langle \Sigma_{\rm g} \rangle$, and
$P_\ell(\cos\theta)$ is the Legendre polynomial.
For small-angle regime, where we are mainly interested, we can
approximate the sky as a flat surface, and then, the above expression
simply becomes two-dimensional Fourier transform:
\begin{equation}
 C_\ell^{\rm dm, g} = \int d^2\theta C_{\rm dm, g}(\theta) e^{-i {\bm
  \ell} \cdot {\bm \theta}}.
\end{equation}
By following discussions in Appendix of Ref.~\cite{Ando2007} (see
also Appendix~\ref{sec:App C} in the present paper), we arrive at the simple
expression:
\begin{equation}
 C_\ell^{\rm dm, g} = \int\frac{d\chi}{\chi^2}W_{\rm dm}(z) W_{\rm g}(z)
  P_{\delta^2, {\rm g}}\left(\frac{\ell}{\chi}, z\right),
\end{equation}
where $P_{\delta^2, {\rm g}}(k, z)$ is the three-dimensional cross-power
spectrum between overdensity squared $\delta^2$ and galaxy distribution
$\delta_{\rm g} = (n_{\rm g} - \langle n_{\rm g} \rangle) / \langle
n_{\rm g} \rangle$.

In the framework of `halo model'~\cite{Scherrer1991, Seljak2000,
Cooray2002}, where one assumes that all the matter including dark matter
as well as galaxies are contained in spherical halos, the cross-power
spectrum is divided into one-halo and two-halo terms:
\begin{equation}
 P_{\delta^2, {\rm g}}(k, z) = P^{\rm 1h}_{\delta^2, {\rm g}}(k, z) +
  P^{\rm 2h}_{\delta^2, {\rm g}}(k, z).
\end{equation}
For the one-halo term, we cross-correlate $\delta^2$ and galaxies in
one single halo, while for the two-halo term, we associate $\delta^2$ in
a halo and $\delta_{\rm g}$ in another distinct halo, and take the cross
correlation between these two.
Therefore, once we know the distributions of both matter and galaxies
within each halo, and also the intrinsic correlation between halo
positions, we are able to compute the cross-correlation power spectrum.
We give detailed derivation in Appendix~\ref{sec:App C}, and here show
the formulae for both the one-halo and two-halo terms:
\begin{eqnarray}
 P_{\delta^2, {\rm g}}^{\rm 1h}(k, z) &=&
  \int dM \frac{dn(M,z)}{dM}
  \frac{\mathcal J(M, z)}{(\Omega_{\rm m} \rho_{\rm c})^2}
  \frac{\langle N_{\rm g}| M \rangle}{\langle n_{\rm g}(z)\rangle}
  \tilde u_{\delta^2}(k|M) \tilde u_{\rm g}(k|M),
  \label{eq:1halo}\\
 P_{\delta^2, {\rm g}}^{\rm 2h}(k, z) &=&
  \left[\int dM \frac{dn(M,z)}{dM}\frac{\mathcal J(M, z)}{(\Omega_{\rm
  m}\rho_{\rm c})^2} \tilde u_{\delta^2}(k|M)
  b_1(M, z)\right]
  \nonumber\\&&{}\times
  \left[\int dM^\prime \frac{dn(M^\prime, z)}{dM^\prime}
  \frac{\langle N_{\rm g}| M^\prime \rangle}{\langle n_{\rm g}(z)\rangle}
   \tilde u_{\rm g}(k|M^\prime) b_1(M^\prime, z)\right] P_{\rm lin}(k,
  z).
  \label{eq:2halo}
\end{eqnarray}
The one-halo term has only one integration over mass function.
It is multiplied by quantities related to matter and galaxy
distributions.
The former is represented by $\mathcal J \tilde u_{\delta^2} /
(\Omega_{\rm m}\rho_{\rm c})^2$ and the latter by $\langle N_{\rm g} | M
\rangle \tilde u_{\rm g} / \langle n_{\rm g} \rangle$, where $\langle
N_{\rm g} | M \rangle$ is the number of galaxies present in a host halo
with mass $M$ and $\langle n_{\rm g} \rangle$ is the average number
density of the galaxies.
See Appendix~\ref{sub:App B1} for more details of these quantities about
the galaxies.
Distribution of density squared $u_{\delta^2}(r)$ and that of the galaxies
$u_{\rm g}(r)$ (both normalized to one after volume integration) are
represented by the Fourier transform of these profiles, $\tilde
u_{\delta^2}(k)$ and $\tilde u_{\rm g}(k)$, respectively.
For the former, we adopt the same model as in Ref.~\cite{AK2013} (see
Eqs. 12--15 there) that is based on numerical simulations of subhalo
distributions~\cite{Gao2012, Han2012}, while for the latter, we assume
that the (satellite) galaxies distribute following the NFW profile (see,
e.g., Ref.~\cite{Cooray2002} for the Fourier transform of the NFW
profile).
The two-halo term, on the other hand, includes two integrations over the
mass function as it depends on contents in two independent halos.
The one integral depends on distribution of $\delta^2$ and the other
does on distribution of the galaxies.
These two factors are connected through the intrinsic correlation
between the two halos, and this halo-halo power spectrum is approximated
by $P_{\rm hh}(k|M, M^\prime) = b_1 (M) b_1(M^\prime) P_{\rm lin}(k)$,
where $b_1(M)$ is the linear bias (e.g., \cite{ST}) and $P_{\rm lin}(k)$
is the linear matter power spectrum.

\begin{figure}[t]
 \begin{center}
  \includegraphics[width=8.5cm]{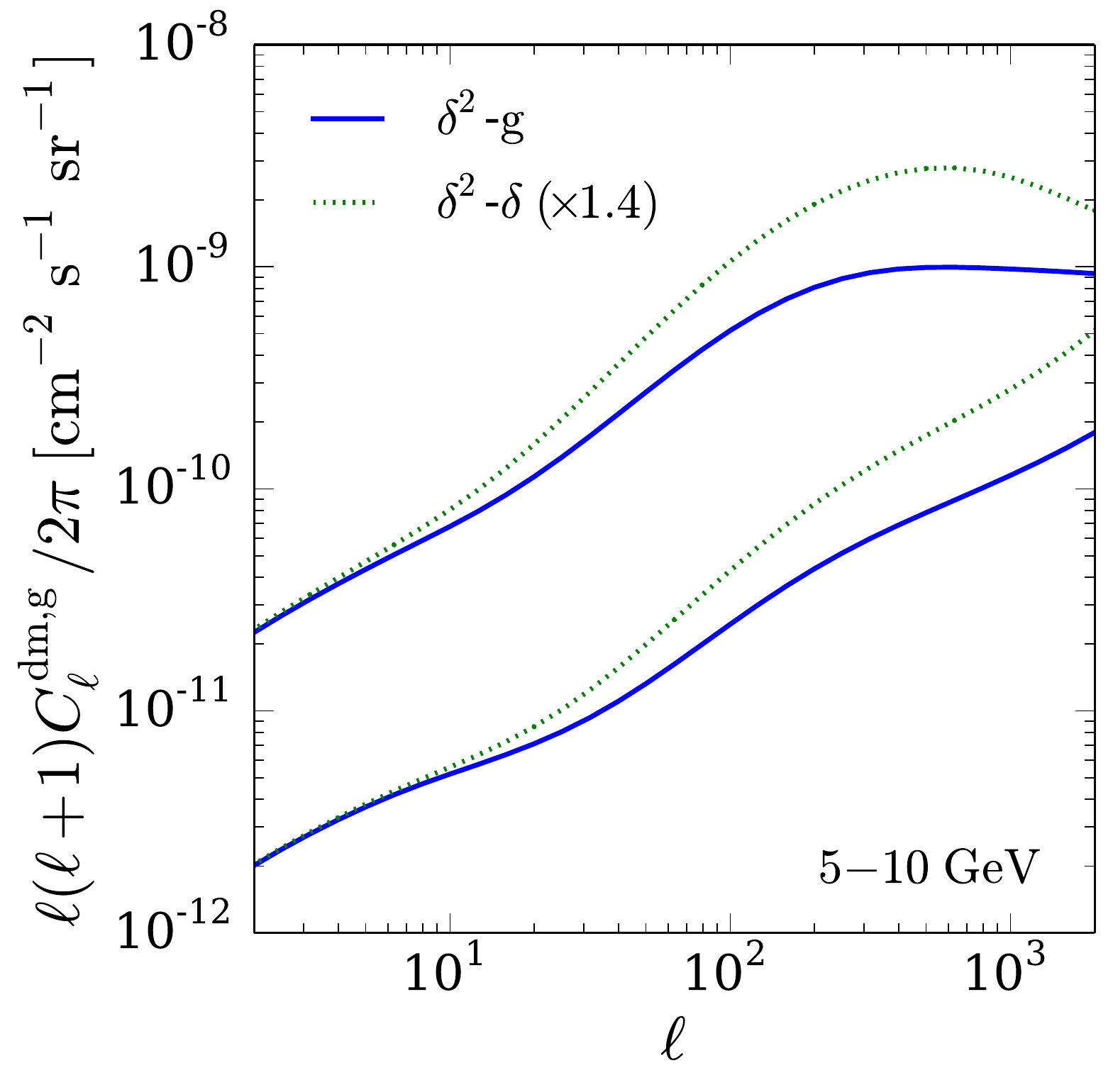}
  \caption{The angular cross-power spectrum between the gamma-ray
  background at 5--10~GeV due to dark matter annihilation and the 2MRS
  galaxies. For dark matter, $m_{\rm dm} = 100$~GeV, $\langle \sigma v
  \rangle = 3 \times 10^{-26}$~cm$^3$~s$^{-1}$, and $b\bar b$
  annihilation channel are assumed. Upper and lower sets of curves are
  for the two boost models~\cite{Gao2012, Sanchez-Conde2014},
  respectively. Solid curves are based on
  the density squared--galaxy power spectrum [$P_{\delta^2, {\rm
  g}}(k)$], while dotted curves are on density squared--density power
  spectrum [$1.4 P_{\delta^2, \delta}(k)$, where 1.4 corrects for galaxy
  bias in the linear regime].}
  \label{fig:Cl_DM_g}
 \end{center}
\end{figure}

In Fig.~\ref{fig:Cl_DM_g}, we show the angular cross-power spectrum
between dark matter annihilation and the 2MRS galaxies for 5--10~GeV
band.
We here compare the resulting spectra from $P_{\delta^2, {\rm g}}(k)$
and those from $P_{\delta^2, \delta}(k)$ multiplied by a constant galaxy
bias at linear regime (assumed to be 1.4).
The $\delta^2$-$\delta$ power spectrum that was also adopted in
Ref.~\cite{ABK} can be evaluated by replacing $\langle N_{\rm g} | M
\rangle \tilde u_{\rm g} / \langle n_{\rm g} \rangle$ with $M \tilde
u_{\delta} / (\Omega_{\rm m}\rho_{\rm c})$ in Eqs.~(\ref{eq:1halo}) and
(\ref{eq:2halo}).
Assuming a constant, scale-independent bias of 1.4 with respect to the
$\delta^2$-$\delta$ power spectrum will then result in an overestimate
of the dark matter cross-power spectrum by up to a factor of a few in
scales most relevant for possible detection.
We also compare the results of the two boost models~\cite{Gao2012,
Sanchez-Conde2014}.
The shapes for these two models are similar except for small angular
scales, but the overall amplitude is larger by about an order of
magnitude for the more optimistic model by Ref.~\cite{Gao2012}.

\begin{figure}[t]
 \begin{center}
  \includegraphics[width=15cm]{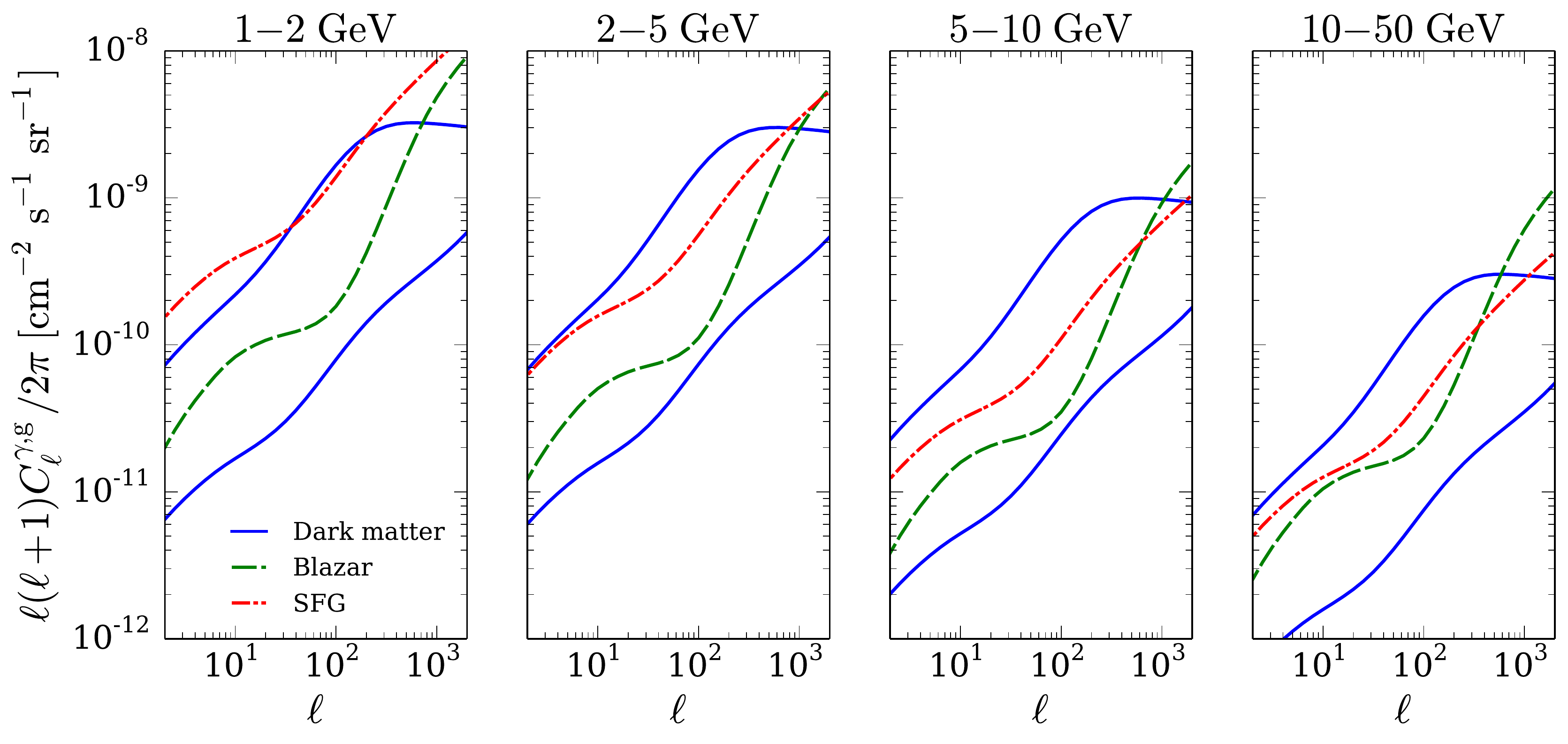}
  \caption{The angular cross-power spectrum between the 2MRS galaxies
  and the gamma-ray background (for four energy bands as shown at the
  top of each panel) due to dark matter, SFGs, and blazars. Dark matter
  components are evaluated for $m_{\rm dm} = 100$~GeV, $\langle \sigma
  v\rangle = 3 \times
  10^{-26}$~cm$^{3}$~s$^{-1}$, pure $b\bar b$ annihilation
  channel, and shown for both the boost factor models~\cite{Gao2012,
  Sanchez-Conde2014}.}
  \label{fig:Cl_energy_2MRS}
 \end{center}
\end{figure}

\begin{figure}[t]
 \begin{center}
  \includegraphics[width=15cm]{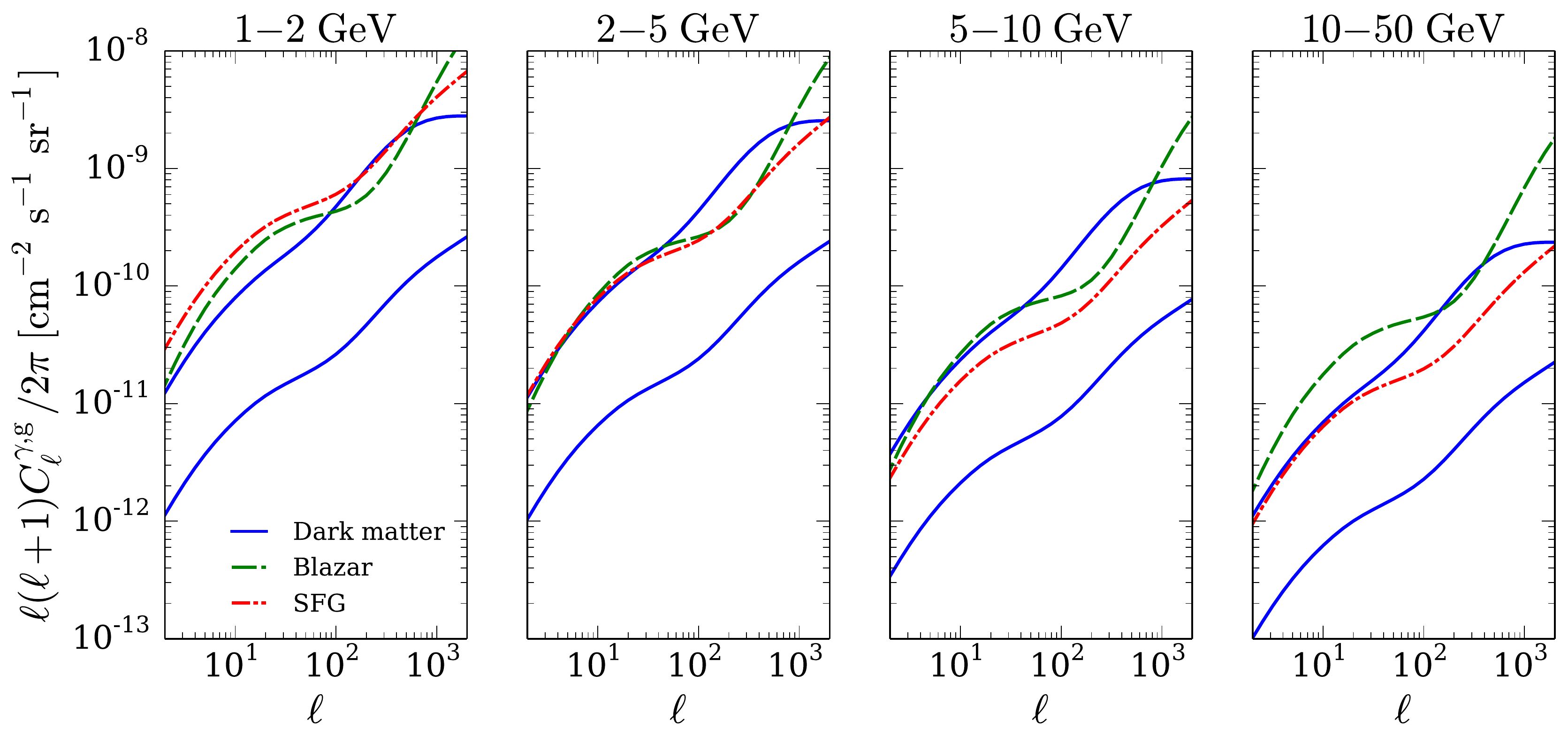}
  \caption{The same as Fig.~\ref{fig:Cl_energy_2MRS} but
  cross-correlated with the 2MXSC galaxies.}
  \label{fig:Cl_energy_2MXSC}
 \end{center}
\end{figure}

Figures~\ref{fig:Cl_energy_2MRS} and \ref{fig:Cl_energy_2MXSC} show the
angular power spectrum $C_\ell^{\rm dm, g}$, cross correlated with the
2MRS and 2MXSC galaxies, respectively, evaluated in the four energy
bins: 1--2, 2--5, 5--10, and 10--50~GeV.
Dark matter components (evaluated again for both the boost
models~\cite{Gao2012, Sanchez-Conde2014}) are compared with
contributions from the astrophysical sources (see Sec.~\ref{sub:Angular
cross-power spectrum: Astrophysical sources}).
One can see that the energy dependence as well as the shape of the power
spectra are characteristic for dark matter, which will be used for
distinguishing that component from the others in the following
discussions.

\subsection{Angular cross-power spectrum: Astrophysical sources}
\label{sub:Angular cross-power spectrum: Astrophysical sources}

The total angular cross-power spectrum is the sum of all the
contributions from astrophysical gamma-ray sources as well as dark
matter annihilation.
Here, as in Sec.~\ref{sub:Astrophysical sources}, we consider the SFGs
and blazars as such representative astrophysical sources, $X = \{{\rm SFG,
b}\}$.
Since most of them are regarded as point-like gamma-ray sources
for Fermi-LAT, and they may overlap with catalog galaxies in 2MRS
or 2MXSC, the cross-power spectrum will also include the stochastic shot
noise that comes from the discrete nature of some sources.
Therefore, the total cross-power spectrum is written as
\begin{equation}
 C_\ell^{\gamma, {\rm g}} = C_\ell^{\rm dm, g} + \sum_X 
  \left(C_\ell^{X, {\rm
  g}} + C_{\rm P}^{X, {\rm g}}\right),
\end{equation}
where $C_{\rm P}^{X, {\rm g}}$ represents the shot (or Poisson) noise of
the source $X$.
The shot-noise term, as we derive in Appendix~\ref{app:shot noise},
depends on the number density of the relevant sources, and results in
being independent of scales or multipoles $\ell$.
Therefore, it is relatively straightforward to subtract this
scale-independent component from the total power spectrum, but such a
procedure induces errors.

The angular cross-power spectrum between the gamma-ray background due to
a source class $X$ and catalog galaxies g, $C_\ell^{X, {\rm g}}$, is
evaluated similarly as $C_\ell^{\rm dm, g}$ in the previous subsection:
\begin{equation}
 C_\ell^{X, {\rm g}} = \int \frac{d\chi}{\chi^2} W_X(z)W_{\rm g}(z)
  P_{X, {\rm g}}\left(\frac{\ell}{\chi}, z\right),
\end{equation}
where $P_{X, {\rm g}}(k, z)$ is the cross-power spectrum between $X$ and
the galaxies (excluding the shot-noise component).
We once again base our approach on the halo model, and give details
in Appendices~\ref{sub:App B1} and \ref{sub:App B2}, for the SFGs and
blazars, respectively.
Here, we simply give qualitative arguments of the model and show results
of the cross-power spectra.

\begin{figure}[t]
 \begin{center}
  \includegraphics[width=8.5cm]{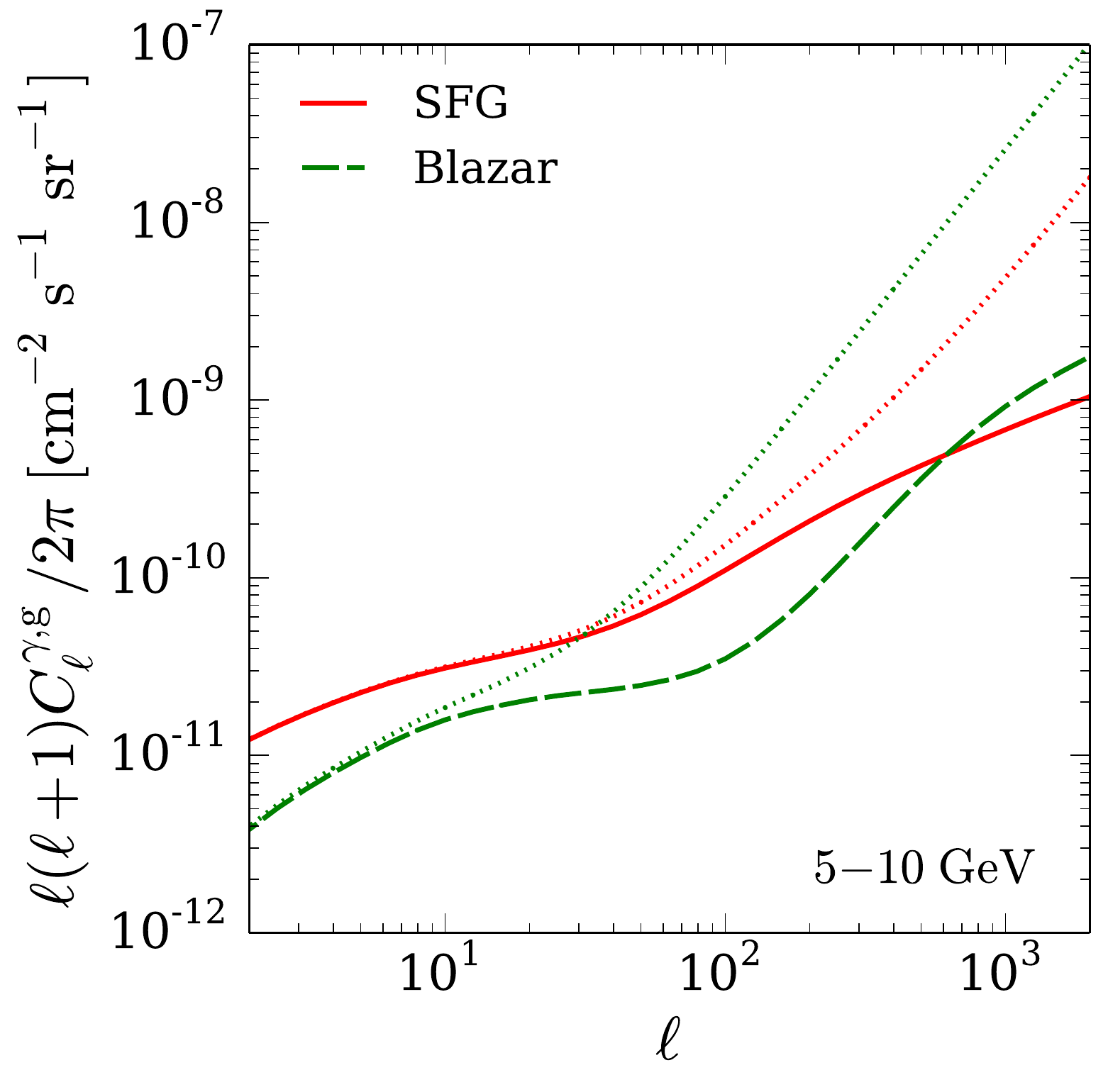}
  \caption{The angular cross-power spectrum between the gamma-ray
  background at 5--10~GeV due to SFGs (solid) or blazars (dashed) and
  the 2MRS galaxies. Dotted curves include the shot-noise terms, $C_{\rm
  P}^{\rm SFG, g}$ and $C_{\rm P}^{\rm b, g}$.}
  \label{fig:Cl_Astro_g}
 \end{center}
\end{figure}

The galaxy power spectrum $P_{\rm g}(k, z)$ is the relevant quantity for
SFGs.
Within the framework of the halo model, this can be evaluated once we
specify a number of galaxies in a halo with mass $M$, $\langle N_{\rm g}
| M \rangle$, and how they are distributed in it.
The angular cross-power spectrum for this component is shown in
Fig.~\ref{fig:Cl_Astro_g} for the 5--10~GeV energy band, and is compared
with the shot noise associated with it.
The shot noise starts dominating at small scales, $\ell \gtrsim 200$,
and thus one cannot ignore such a term.

Clustering properties of the gamma-ray blazars are much less understood,
mainly because of paucity of detected sources.
We here assume that the gamma-ray blazars are well correlated with
AGNs identified with X rays.
In fact, the blazar luminosity functions adopted here~\cite{Ajello2012,
Ajello2014} are constructed relying on the same argument.
By studying the distribution of X-ray AGNs, Refs.~\cite{Allevato2011,
Cappelluti2012} found that they appear to selectively locate in dark
matter halos with mass around $\sim$10$^{13.1}\, h^{-1} M_\odot$.
Following this, we assume that the gamma-ray blazars also live in these
halos (at the center), whose bias parameter is then obtained as the one
for their host halos, and also that there is no more than one blazar in
each halo.
We show the angular power spectrum for blazars calculated this way
in Fig.~\ref{fig:Cl_Astro_g}, cross-correlated with the 2MRS galaxies,
as well as the shot-noise component.
Since fewer and brighter blazars contribute to the gamma-ray background
compared with SFGs, they yield larger shot-noise term, which becomes
dominant already at $\ell \sim 30$.

Figures~\ref{fig:Cl_energy_2MRS} and \ref{fig:Cl_energy_2MXSC} summarize
the energy dependence of both the blazar and SFG components,
cross-correlated with the 2MRS and 2MXSC galaxies, respectively.

\section{Sensitivity to annihilation cross section}
\label{sec:sensitivity}

\subsection{Covariance matrix and errors of cross correlation}
\label{sub:errors}

We consider gamma-ray maps $\gamma_i, \gamma_j, \cdots$ and galaxy
catalogs ${\rm g}_a, {\rm g}_b, \cdots$, where one can think that $i, j,
\cdots$ represent energy bins, and $a, b, \cdots$ do redshift bins.
If we consider one catalog (such as 2MRS or 2MXSC) as a whole, then
there is only one redshift bin to be considered.
Covariance between $C_\ell^{\gamma_i, {\rm g}_a}$ and $C_\ell^{\gamma_j,
{\rm g}_b}$ is given by (e.g., \cite{Tristram2005, Cuoco2011})
\begin{equation}
 \mbox{Cov}\left(C_\ell^{{\gamma_i}, {\rm g}_a}, C_\ell^{\gamma_j, {\rm
  g}_b}\right) = \frac{1}{(2\ell + 1)f_{\rm sky}}
  \left[ C_\ell^{\gamma_i, {\rm g}_b} C_\ell^{\gamma_j, {\rm g}_a}
   + \left(C_\ell^{\gamma_i, \gamma_j} + \delta_{ij}
      \frac{C_{\rm N}^{\gamma_i}}{W_\ell^{i 2}}
     \right)
  \left(C_\ell^{{\rm g}_a, {\rm g}_b} + \delta_{ab} C_{\rm N}^{{\rm
   g}_a}\right)\right],
  \label{eq:cov}
\end{equation}
where $f_{\rm sky}$ is the fraction of sky covered by the survey,
$\delta_{ij}$ and $\delta_{ab}$ are the Kronecker delta, $C_{\rm
N}^{\gamma_i}$ and $C_{\rm N}^{{\rm g}_a}$ are the shot noise of the
photons in the energy bin $i$ and that of the galaxies in the redshift
bin $a$, respectively, and $W_\ell^i$ is the Fourier transform of the
point-spread function in the energy band $i$.

\begin{figure}[t]
 \begin{center}
  \includegraphics[width=12cm]{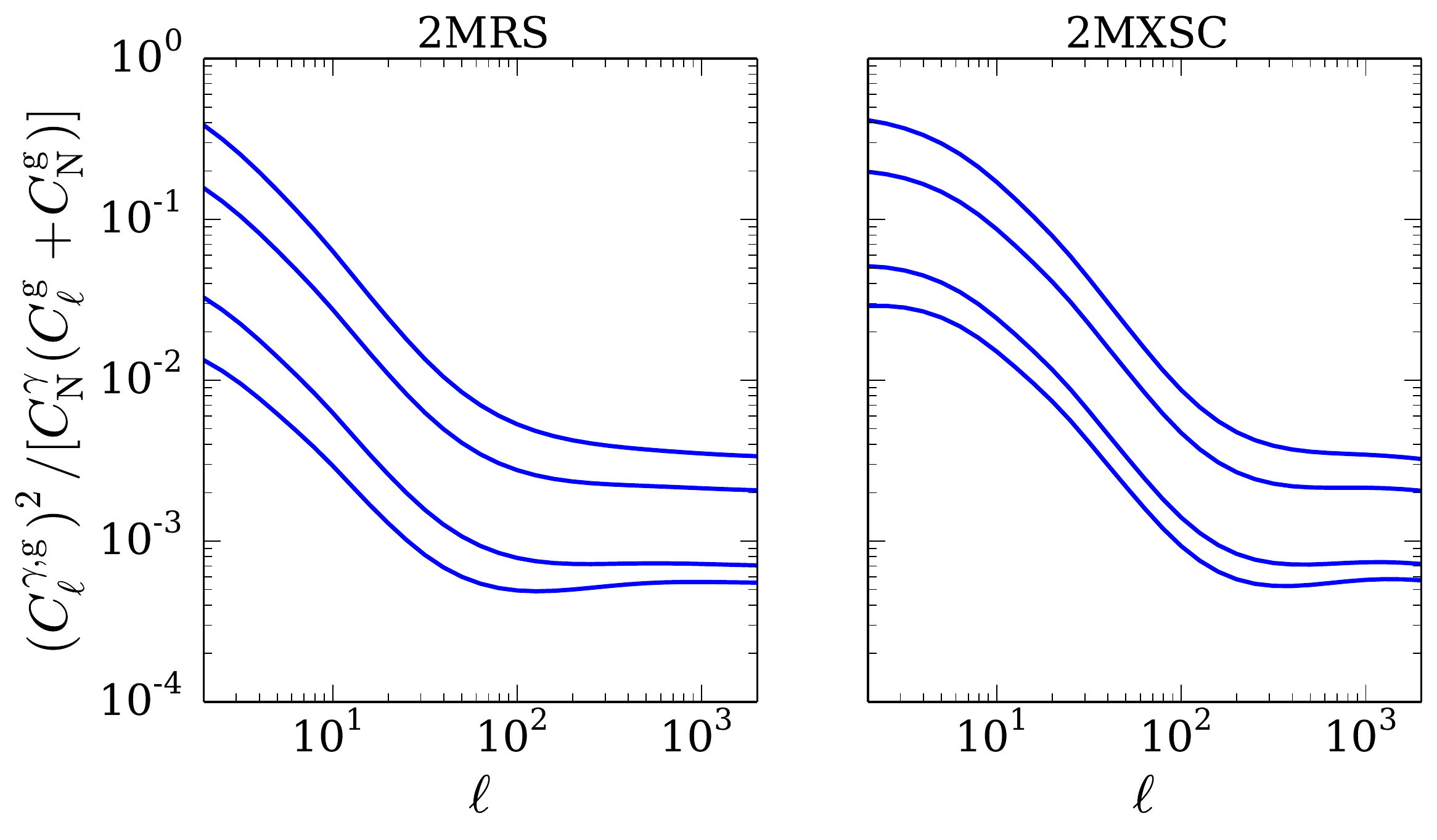}
  \caption{The ratio between cross-power spectrum squared and product of
  gamma-ray and galaxy auto-power spectra, $(C_\ell^{\gamma, {\rm g}})^2
  / [C_{\rm N}^\gamma (C_\ell^{\rm g} + C_{\rm N}^{\rm g})]$, for the
  2MRS (left) and 2MXSC (right) galaxies. Four curves in each panel
  correspond to the energy bands, 1--2, 2--5, 5--10, and 10--50~GeV
  (from top to bottom). A plateau for $\ell \gtrsim 100$ is due to shot
  noise in the cross power.}
  \label{fig:covariance}
 \end{center}
\end{figure}

If two redshift bins do not overlap, as we postulate in this
paper, then there is no cross correlation in the galaxy power spectrum;
the galaxies in different redshift ranges are uncorrelated, i.e.,
$C_\ell^{{\rm g}_a, {\rm g}_b} = \delta_{ab} C_\ell^{{\rm g}_a}$ (see
also discussions at the end of this subsection).
One cannot argue in a similar way for $C_\ell^{\gamma_i, \gamma_j}$,
because a bright source in one energy bin $i$ tends to be also bright in
another energy bin $j$.
In practice, however, the photon shot noise $C_{\rm N}^\gamma$ is much
larger than the angular auto-power spectrum $C_\ell^\gamma$ from the
anisotropy measurement with Fermi~\cite{FermiAnisotropy}, and therefore
one can treat the second term in the square bracket of
Eq.~(\ref{eq:cov}) as a diagonal matrix proportional to $\delta_{ij}
\delta_{ab}$.
We can further show that the first term is negligible compared with the
second, by explicitly computing $(C_\ell^{\gamma, {\rm g}})^2$ and
comparing it with $C_{\rm N}^\gamma (C_{\ell}^{\rm g} + C_{\rm N}^{\rm
g})$.
Figure~\ref{fig:covariance} shows $(C_\ell^{\gamma, {\rm g}})^2 /
[C_{\rm N}^\gamma (C_\ell^{\rm g} + C_{\rm N}^{\rm g})]$ for both the
2MRS and 2MXSC galaxies 
and various energy bins, which is indeed found to be smaller than one in
any of these cases.
If we divide these catalogs into redshift bins, then the galaxy
shot noise $C_{\rm N}^{{\rm g}}$ increases, yielding even smaller values
for this ratio.
Therefore, we can safely approximate Eq.~(\ref{eq:cov}) as a diagonal
matrix:
\begin{equation}
 \mbox{Cov}\left(C_\ell^{\gamma_i, {\rm g}_a}, C_\ell^{\gamma_j, {\rm
	    g}_b}\right)
  \approx \delta_{ij}\delta_{ab}
  \left(\delta C_\ell^{\gamma_i, {\rm g}_a} \right)^2,
  \label{eq:diagonal}
\end{equation}
 where the diagonal components are given as
\begin{equation}
 \left(\delta C_\ell^{\gamma_i, {\rm g}_a}\right)^2 = \frac{1}{(2\ell +
  1) f_{\rm sky}}
  \left[\left(C_\ell^{\gamma_i, {\rm g}_a}\right)^2 + 
  \left(C_\ell^{\gamma_i} + \frac{C_{\rm
   N}^{\gamma_i}}{W_\ell^{i2}}\right)
  \left(C_\ell^{{\rm g}_a} + C_{\rm N}^{{\rm g}_a}\right)\right].
  \label{eq:error}
\end{equation}

For the sensitivity study below, we adopt $f_{\rm sky} = 0.7$, $C_{\rm
N}^{\gamma_i} = I_{\rm obs}^i / \mathcal E$, $C_{\rm N}^{{\rm g}_a} = 4
\pi f_{\rm g} / N_{{\rm g}_a}$, where $I_{\rm obs}^i$ is the observed
mean intensity in the energy band $i$ reported in
Ref.~\cite{FermiDiffuse}, $\mathcal E = 1.5 \times 10^{11}$~cm$^2$~s is
five-year exposure of Fermi-LAT, $N_{{\rm g}_a}$ is the number of
the catalog galaxies contained in redshift bin $a$ from $4\pi f_{\rm g}$
sr of the sky; $N_{\rm g} = 43500$ (770000) and $f_{\rm g} = 0.91$
(0.67) for 2MRS (2MXSC) without redshift binning.
The angular auto-power spectrum of the gamma-ray background
$C_\ell^{\gamma_i}$ is found to be dominated by shot-noise due to the
blazars~\cite{FermiAnisotropy, Cuoco2012}, and we adopt these measured
values.
For the angular response represented by $W_\ell^i$, we use the results
reported in Ref.~\cite{FermiAnisotropy}.
We also note that $C_\ell^{\gamma_i, {\rm g}_a}$ in the right-hand side
of Eq.~(\ref{eq:error}) includes the shot-noise component.

\begin{figure}[t]
 \begin{center}
  \includegraphics[width=8.5cm]{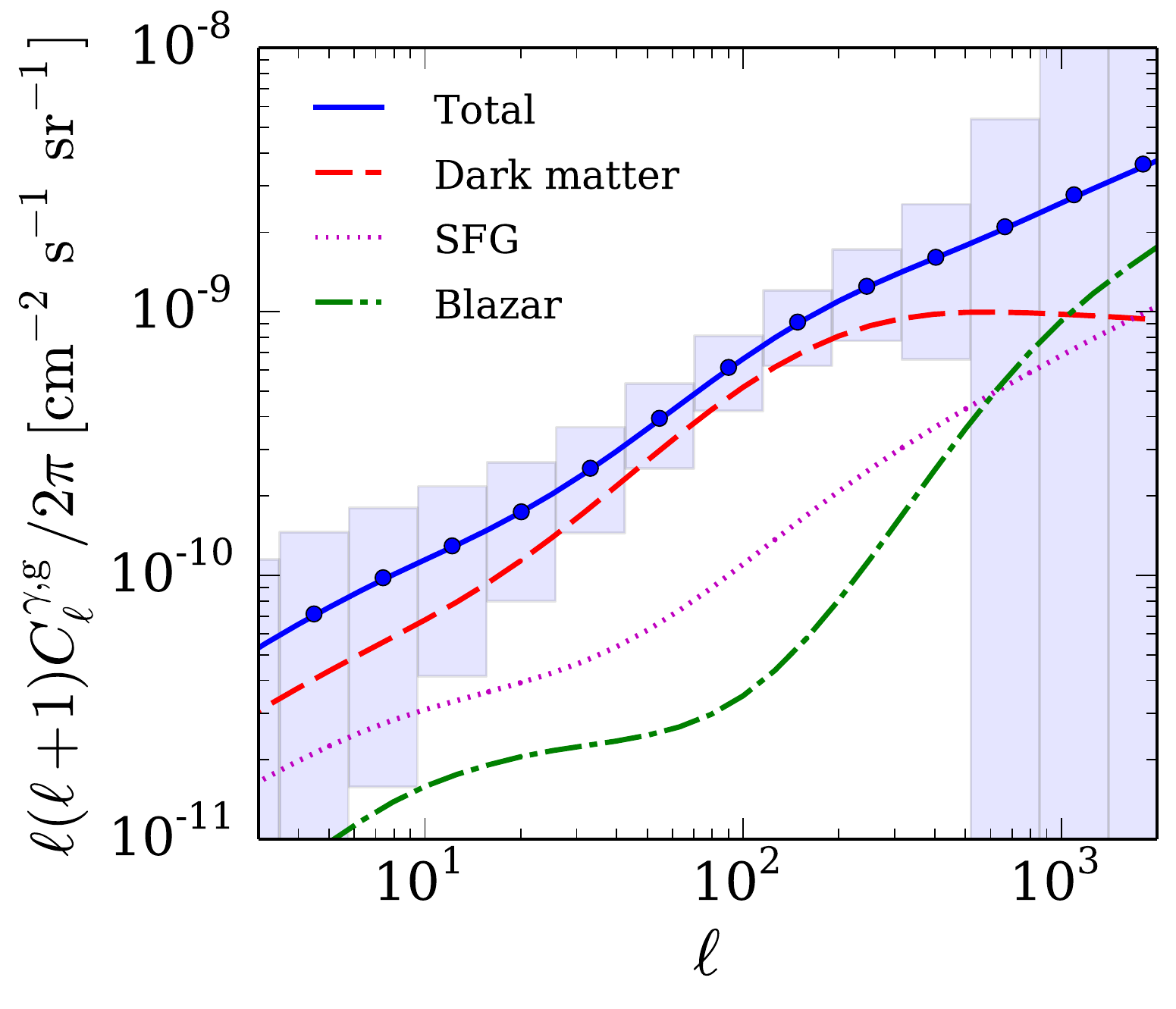}
  \caption{The angular cross-power spectrum between the gamma-ray
  background in the 5--10~GeV band and the 2MRS galaxies, and expected $1\sigma$
  errors (boxes). Dark matter component is evaluated for $m_{\rm dm} =
  100$~GeV, $\langle \sigma v\rangle = 3 \times
  10^{-26}$~cm$^{3}$~s$^{-1}$, pure $b\bar b$ annihilation channel, and
  for the boost model by Ref.~\cite{Gao2012}.}
  \label{fig:Cl_error}
 \end{center}
\end{figure}

As an example of error estimates, Fig.~\ref{fig:Cl_error} shows the
expected errors for the total
cross-power spectrum (after subtracting the shot-noise component) between
the gamma-ray background in the 5--10 GeV energy band and the 2MRS galaxies,
for the boost model by Ref.~\cite{Gao2012}.
The expected errors are very small compared with the signal, and
therefore, if this were the case, the dark matter component would be
clearly detected.
We also note that the prospect is even better than the one from our
previous study.
One can see this by comparing Fig.~\ref{fig:Cl_error} with Fig.~2 of
Ref.~\cite{ABK}.
This comes from difference of the models adopted; here we adopt the
cross-power spectrum based on the HOD of galaxies within halo model,
whereas in Ref.~\cite{ABK}, we simply assumed that galaxies trace matter
with a constant bias.
As the result of this difference, our present model yields larger cross
power, and hence smaller relative errors.

\begin{table}[t]
 \begin{center}
 \caption{Galaxy catalogs for cross correlation. The boundaries of the
  redshift binning and the number of galaxies per bin are shown in the
  second and the third columns, respectively.}
 \label{table:catalog}
 \begin{tabular}[t]{lrr} \hline
  Catalog & Redshift boundaries & $N_{\rm g}$ per bin \\ \hline
  2MRS & (0.003, 0.1) & 43500 \\
  2MRS-N2 & (0.003, 0.027, 0.1) & 21750\\
  2MRS-N3 & (0.003, 0.021, 0.035, 0.1) & 14500\\
  \hline
  2MXSC & (0.003, 0.3) & 770000 \\
  2MXSC-N2 & (0.003, 0.083, 0.3) & 385000\\
  2MXSC-N3 & (0.003, 0.066, 0.10, 0.3) & 257000\\
  2MXSC-N4 & (0.003, 0.058, 0.083, 0.11, 0.3) & 193000\\
  2MXSC-N5 & (0.003, 0.052, 0.073, 0.093, 0.12, 0.3) & 154000 \\
  2MXSC-N10 & (0.003, 0.039, 0.052, 0.063, 0.073,  &  77000 \\
   & 0.083, 0.093, 0.10, 0.12, 0.14, 0.3) & \\\hline
 \end{tabular}
 \end{center}
\end{table}

The redshift binning is simply done such that each bin contains the same
number of galaxies.
Table~\ref{table:catalog} summarizes the catalogs and the redshift
binning.
For example, 2MXSC-N5 is based on the 2MXSC catalog, but we
divide it into five redshift bins, each of which contains 154000
galaxies.
When we use a finer redshift binning, more information is made available.
However, one starts to see correlation between two neighboring redshift
bins, if their widths are too small.
This may be estimated by comparing the width with the galaxy correlation
length---the length at which the two-point correlation function
$\xi_{\rm g}$ becomes one.
We find this length to be $\sim$10~Mpc, by computing $\xi_{\rm g}$ from
the Fourier transform of the galaxy power spectrum $P_{\rm g}(k)$ in the
local Universe ($z = 0$).
The finest redshift slices are obtained for 2MXSC-N10, where $(\Delta
z)_{\rm min} = 0.01$ and this corresponds to the comoving distance of
$\sim$40~Mpc.
This is reasonably larger than the galaxy correlation length, and in
fact the galaxy two-point correlation function at the separation of
40~Mpc is only 0.09.

Another important quantity to compare the width of redshift bins with is
the accuracy of the photometric-redshift determination, in the case of
2MXSC.
The latest analysis of the 2MASS photometric redshift catalog shows that
the redshift accuracy is about 12\%~\cite{Bilicki2014}.
For typical 2MXSC galaxies around $z \sim 0.1$, this corresponds to the
accuracy of $\sigma_z \sim 0.01$, which is comparable to the smallest
bin width for 2MXSC-N10.
Therefore, the redshift slicing using more than several bins might
induce correlation between neighboring redshift bins.
As we shall show in Sec.~\ref{sec:sensitivity}, however, the sensitivity
to the annihilation cross section saturates when using a few redshift
bins already, and therefore, this uncertainty results in no major impact
on our conclusions.

\subsection{Bayesian statistics and prior distributions}
\label{sub:prior}

We adopt the Bayesian statistics (e.g., \cite{Trotta2008}) in order to
obtain the sensitivity to the annihilation cross section.
Given data $\{\bm d\}$, the {\it posterior} probability distribution
function of parameters $\{ \bm \vartheta \}$ is constructed as
\begin{equation}
 P({\bm \vartheta} | {\bm d} ) \propto P({\bm \vartheta})
  P({\bm d} | {\bm \vartheta}),
\end{equation}
where $P(\bm\vartheta)$ is the {\it prior} of the parameters and
$P({\bm d} | {\bm \vartheta})$ is the likelihood function---the
probability of obtaining the data $\{\bm d\}$ given the parameters
$\{\bm \vartheta\}$.
In general, functional form of the posterior is undefined, and in
such a case, a powerful and efficient approach to estimate it is to
perform the MCMC.

The data $\{\bm d\}$ are obtained only through relevant analysis of the
gamma-ray maps cross-correlated with galaxy catalogs.
It is, however, beyond the scope of this paper, as our aim is to
evaluate potential sensitivities of the cross-correlation analysis to
dark matter parameters, especially the annihilation cross section.
Therefore, for the current purpose, it suffices to construct `data' from
the model itself.
The model is defined with fixed values of dark matter mass $m_{\rm dm}$
and shapes of the cross-power spectra.
As parameters \{$\bm\vartheta$\} that vary, we consider the following
four: the annihilation cross section $\langle \sigma v \rangle$, the
amplitudes of the cross-power spectra for blazars $\vartheta_{\rm b}$
and for SFGs $\vartheta_{\rm SFG}$, and amplitude of the shot-noise
component $\vartheta_{\rm P}$.
The data of the cross-power spectra are assumed to be the same as a
model, with no dark matter component ($\langle \sigma v
\rangle = 0$), but assuming that both the blazar and SFG contributions
are described by the reference models given in Sec.~\ref{sub:Angular
cross-power spectrum: Astrophysical sources}, and we normalize the
parameters to  $\vartheta_{\rm b} = 1$, $\vartheta_{\rm SFG} = 1$,
and $\vartheta_{\rm P} = 1$ in this reference case:
\begin{equation}
 {\bm d} \equiv \left(C_\ell^{\gamma_i, {\rm g}_a}\right)_{\rm data} =
  \sum_{X = \{{\rm b, SFG}\}}
  \left[\left(C_\ell^{X_i, {\rm g}_a}\right)_{\rm ref} + \left(C_{\rm
	 P}^{X_i, {\rm g}_a}\right)_{\rm ref}\right],
  \label{eq:data}
\end{equation}
where the subscript `ref' stands for the reference theoretical models.

The likelihood function is then obtained as the product of (assumed)
Gaussian distribution of the cross-power spectrum as follows:
\begin{eqnarray}
 P(\bm d | \bm \vartheta) & \propto & \frac{1}{(\det \mbox{Cov})^{1/2}}
  \exp\left\{-\frac{1}{2}
       \left[{\bm d} - {\bm C}(\bm\vartheta )\right]^T
  \mbox{Cov}\left[{\bm C}(\bm\vartheta), {\bm C}(\bm\vartheta)\right]^{-1}
  \left[{\bm d} - {\bm C}(\bm\vartheta )\right]	\right\}
  \nonumber\\&\approx&
  \prod_i\prod_a\prod_\ell\frac{1}{\delta C_\ell^{\gamma_i, {\rm
		    g}_a}(\bm\vartheta)}
  \exp\left\{-\frac{1}{2}
  \left[\frac{\left(C_\ell^{\gamma_i, {\rm g}_a}\right)_{\rm data} -
   C_\ell^{\gamma_i, {\rm
   g}_a}\left(\bm\vartheta\right)}{
  \delta C_\ell^{\gamma_i, {\rm g}_a}(\bm\vartheta)}\right]^2 \right\},
  \nonumber\\
  \label{eq:likelihood}\\
 \bm C(\bm\vartheta) &\equiv& C_\ell^{\gamma_i, {\rm g}_a}(\bm\vartheta)
  =
  \frac{\langle\sigma v \rangle}{\langle \sigma v\rangle_{\rm
   ref}} \left(C_\ell^{{\rm dm}_i, {\rm g}_a}\right)_{\rm ref}
   + \vartheta_{\rm b}
   \left(C_\ell^{{\rm b}_i, {\rm g}_a}\right)_{\rm ref}
   + \vartheta_{\rm SFG}
   \left(C_\ell^{{\rm SFG}_i, {\rm g}_a}\right)_{\rm ref}
   \nonumber\\&& {}
   + \vartheta_{\rm P}
   \left[
   \left(C_{\rm P}^{{\rm b}_i, {\rm g}_a}\right)_{\rm ref} +
   \left(C_{\rm P}^{{\rm SFG}_i, {\rm g}_a}\right)_{\rm ref}
   \right],
\end{eqnarray}
where $\langle \sigma v \rangle_{\rm ref} = 3 \times
10^{26}$~cm$^{3}$~s$^{-1}$.
In the second equality of Eq.~(\ref{eq:likelihood}), we
approximated the covariance matrix as diagonal
[Eq.~(\ref{eq:diagonal})], and in the denominator of the last expression
of the same equation, we use Eq.~(\ref{eq:error}) but with
$C_\ell^{\gamma_i, {\rm g}_a}(\bm\vartheta)$ in its right-hand side.

For the prior $P(\bm \vartheta)$, we assume flat distributions in
logarithmic space for the following ranges: $-30 < \log (\langle \sigma
v \rangle / \mathrm{cm^3~s^{-1}}) < -20$, $-5 < \log\vartheta_{\rm b} <
2$, $-5 < \log\vartheta_{\rm SFG} < 2$, and $-2 < \log\vartheta_{\rm P}
< 2$.
These cover sufficiently large volume in the parameter space.
This is a conservative approach, as we do not use information from any
other measurements such as amplitudes for luminosity functions of SFGs
and blazars.

\subsection{Markov-Chain Monte Carlo simulations and sensitivity estimates}
\label{sub:MCMC}

\begin{figure}[t]
 \begin{center}
  \includegraphics[width=15cm]{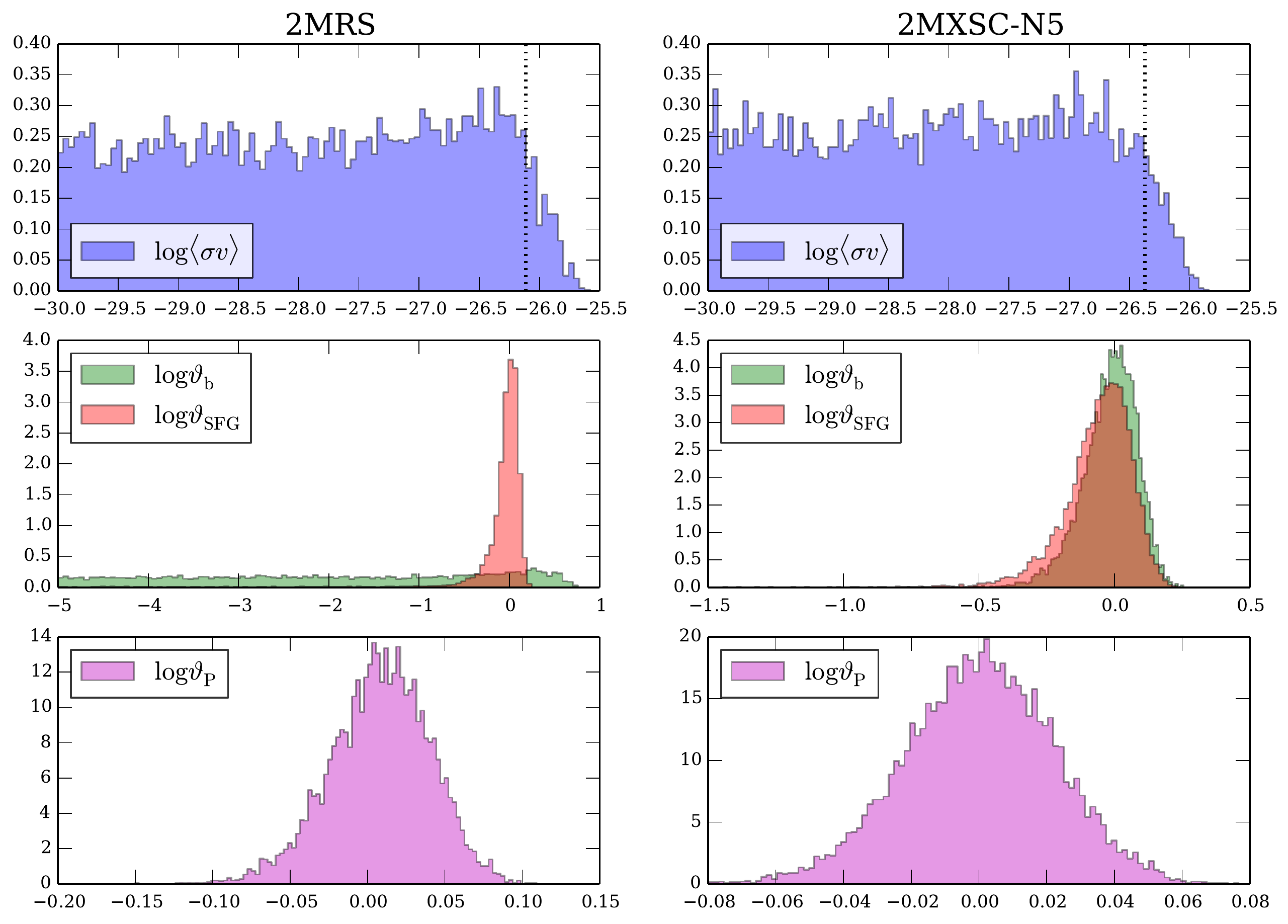}
  \caption{Posterior distributions of parameters ($\langle \sigma v
  \rangle$, $\vartheta_{\rm b}$, $\vartheta_{\rm SFG}$, and
  $\vartheta_{\rm P}$) for 2MRS (left) and 2MXSC-N5 (right). The dark
  matter mass is 100~GeV and the boost model is by
  Ref.~\cite{Gao2012}. The dotted vertical lines in the top panels
  indicate the 95\% credible upper limit.}
  \label{fig:posterior}
 \end{center}
\end{figure}

We obtain the posterior distributions of the parameters
$\{\bm\vartheta\}$ by running MCMC.
Figure~\ref{fig:posterior} shows them for 2MRS (left) and 2MXSC-N5
(right), in the case of the 100-GeV dark matter
annihilating purely into $b\bar b$ and the optimistic boost
model~\cite{Gao2012}.
First, the shot-noise component is well determined from its
scale-independent feature, in particular important at high multipoles.
The posterior distribution of $\vartheta_{\rm P}$ has width of only a
few to several percent (bottom panels).
The SFGs can also be determined relatively well, but the blazar
component is unconstrained by using the 2MRS catalog (middle left
panel).
This is due to limited redshift information available for the
2MRS catalog, and therefore, can be improved by using more than one
redshift bin as shown, e.g., in the case of 2MXSC-N5 (middle right
panel).
Since the `data' do not contain any dark matter component, we have the
posterior distribution of $\langle \sigma v \rangle$ featuring tail at
low values.
Too large values, however, are not allowed as they become seriously in
conflict with the data; hence the distribution features upper cutoff
(top panels).
(The same argument applies to $\vartheta_{\rm b}$ in the case of 2MRS.)
By integrating this posterior distribution up to some value such that
the integral yields 95\% of the total area, one can set the 95\%
credible upper limit.
The upper limit computed this way is shown as dotted vertical lines in
the top panels.

\begin{figure}[t]
 \begin{center}
  \includegraphics[width=8.5cm]{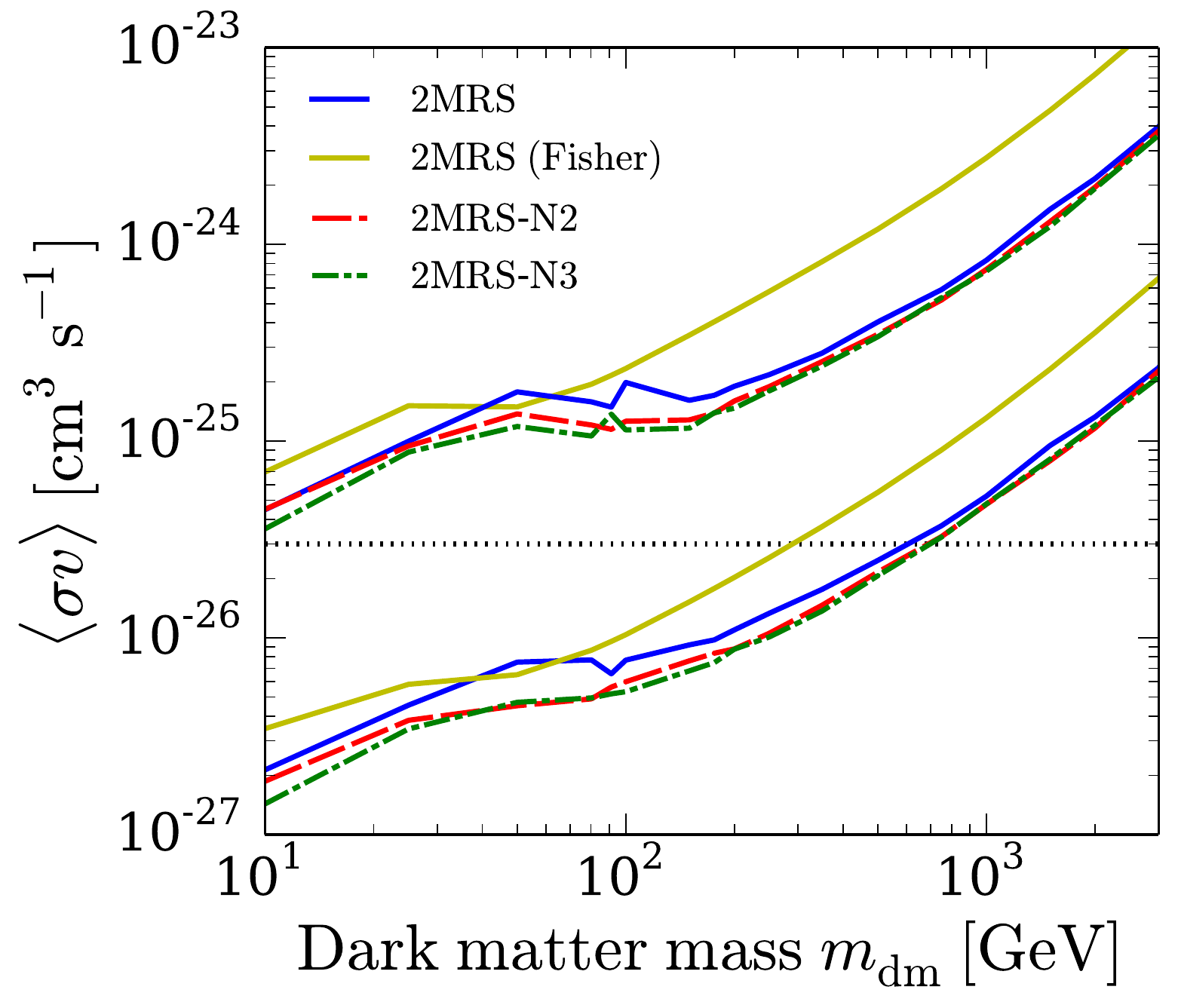}
  \caption{95\% credible sensitivities to the annihilation cross section
  as a function of dark matter mass, obtained from cross correlation
  between the five-year Fermi data for the gamma-ray background and the
  galaxy catalogs (2MRS, 2MRS-N2, and 2MRS-N3). Sensitivities based on
  Fisher matrix as in Ref.~\cite{ABK} are also shown for
  comparison. Lower and upper sets of the curves correspond to
  the results for the different boost models~\cite{Gao2012,
  Sanchez-Conde2014}, respectively. The dotted horizontal line
  represents the canonical annihilation cross section for thermal
  production in the early Universe.}
  \label{fig:sigmav_2MRS}
 \end{center}
\end{figure}

We performed the MCMC described above for all the galaxy catalogs
summarized in Table~\ref{table:catalog} and for various dark matter
masses.
In Fig.~\ref{fig:sigmav_2MRS}, we show 95\% credible sensitivities to
the annihilation cross section as a function of dark matter mass, for
the different boost models~\cite{Gao2012, Sanchez-Conde2014}.
We also show the upper limits for 2MRS using the Fisher matrix,
calculated as in Ref.~\cite{ABK} by varying all the four parameters.
First, we see that the five-year sensitivity for the 2MRS catalog
obtained with MCMC is better than the estimate obtained with the
Fisher matrix, especially at high-mass regime.
In particular, for the optimistic boost case~\cite{Gao2012}, one would
be able to exclude the canonical annihilation cross section for dark
matter less massive than $\sim$700~GeV.
This difference comes from the fact that the method relying on MCMC is
capable of adopting the prior distribution, hence avoiding unphysical
parameter ranges such as negative values for the amplitudes, etc.
Second, by subdividing the 2MRS catalog into a few redshift bins, one
can further improve the sensitivity by up to a factor of two.
The improvement appears to saturate already for the 2MRS-N2 model.
Third, even for the conservative boost
scenario~\cite{Sanchez-Conde2014}, the sensitivity to the cross section
is encouragingly close to its canonical value for thermal production of
dark matter particles.
It features a plateau from tens to a few hundreds of GeV around
$\sim$10$^{-25}$~cm$^{3}$~s$^{-1}$.

\begin{figure}[t]
 \begin{center}
  \includegraphics[width=8.5cm]{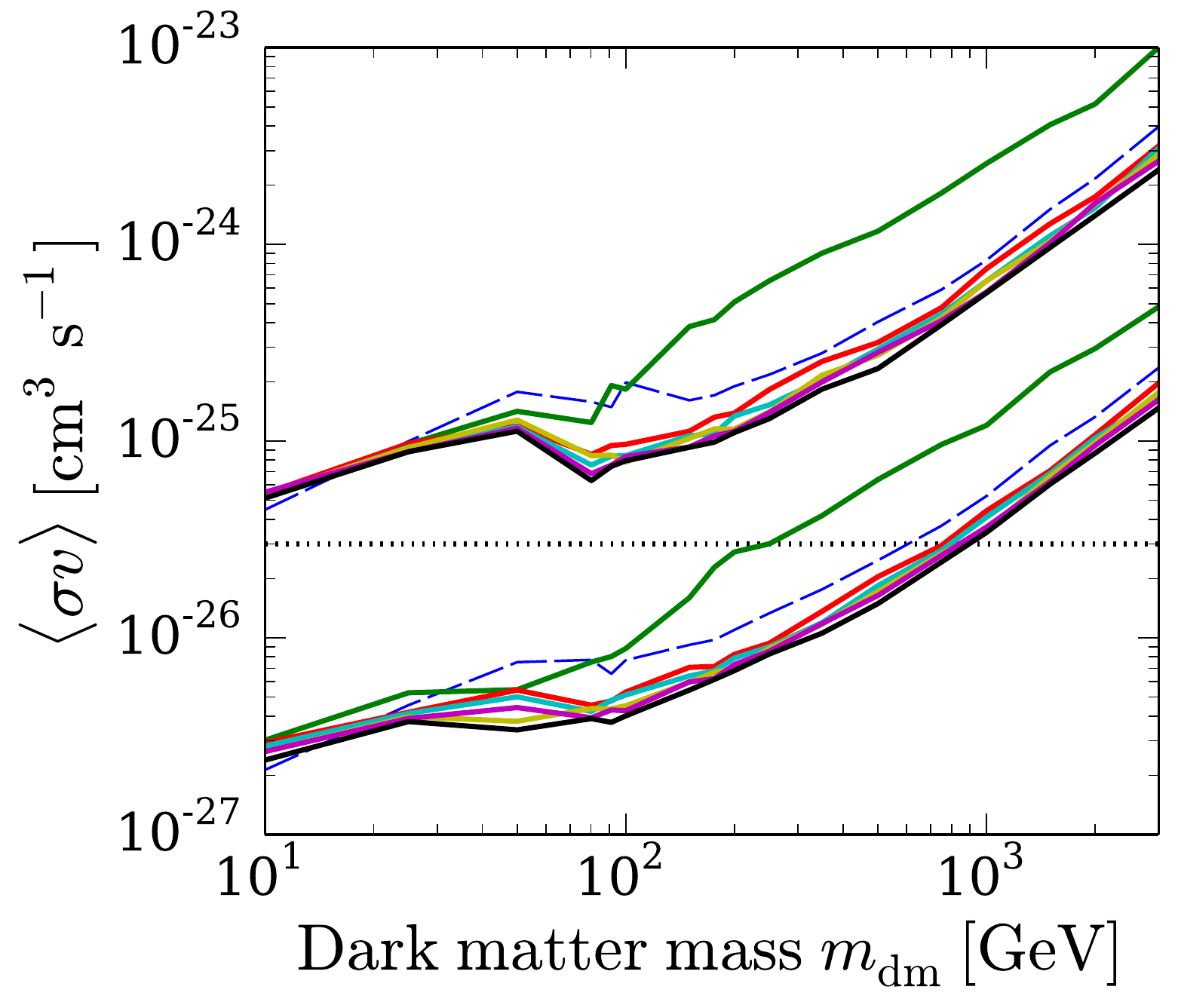}
  \caption{The same as Fig.~\ref{fig:sigmav_2MRS} but for 2MXSC (top
  solid) and for 2MXSC-N2 to 2MXSC-N10 (bottom solid curves). The
  results for 2MRS are shown for comparison as dashed curves.}
  \label{fig:sigmav_2MXSC}
 \end{center}
\end{figure}

Figure~\ref{fig:sigmav_2MXSC} shows the same sensitivity to the
annihilation cross section as Fig.~\ref{fig:sigmav_2MRS}, but for the
larger galaxy catalog 2MXSC.
Even though it is larger, more astrophysical contamination from the
high-redshift regime will weaken the sensitivity compared to the case
with 2MRS.
The tomographic approach, however, proves to be much more efficient in
this case.
If we divide the 2MXSC catalog into two redshift bins (2MXSC-N2), the
expected sensitivity improves already by a factor of several for the
mass range larger than 100~GeV.
There are no further major improvements even if we sub-divide the
catalog into finer redshift slices.
We show all the results of the sensitivity from 2MXSC-N2 to 2MXSC-N10.
Compared with the 2MRS limits, one could improve up to a factor of
three.

Until this point, we assumed that the data are perfectly the same as our
reference model without dark matter [Eq.~(\ref{eq:data})], and
correspondingly obtained the sensitivities to the annihilation cross
section.
In reality, however, data fluctuate statistically, and as the result,
the associated sensitivity estimates also do.
In order to estimate significance of such fluctuations, we generated
data from Monte Carlo simulation.
More specifically, for one Monte Carlo realization, we simulate data
from the Gaussian distribution with the mean of Eq.~(\ref{eq:data}) and
variance of Eq.~(\ref{eq:error}) (again assuming no dark matter
component).
The 95\% credible sensitivities to the annihilation cross section for
this data set were then obtained with the same MCMC procedure as above.
We repeated such a procedure a number of times to obtain the
distribution of the expected sensitivities.

\begin{figure}[t]
 \begin{center}
  \includegraphics[width=8.5cm]{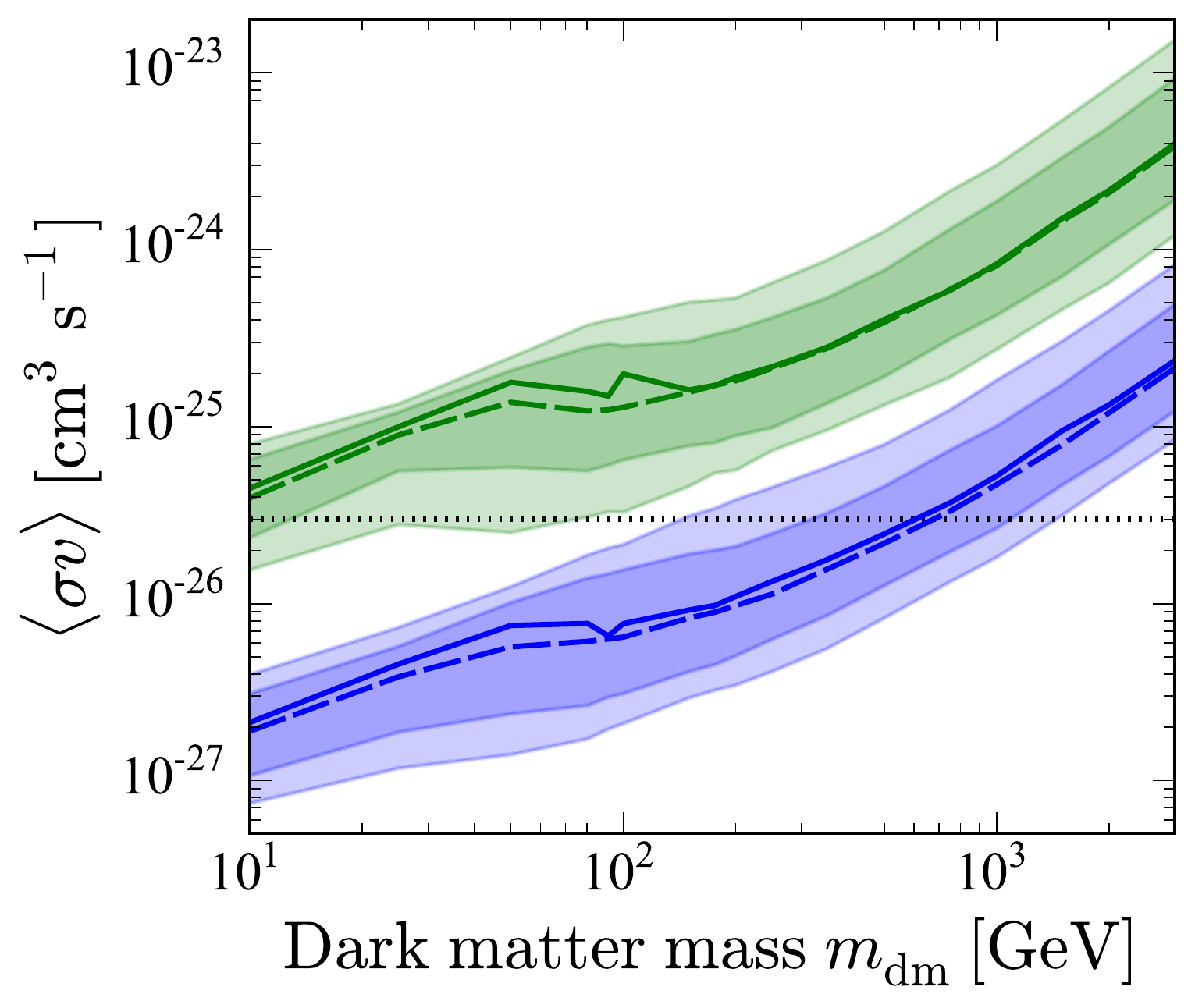}
  \caption{The 68\% and 95\% containment intervals of the 95\% credible
  sensitivity to the annihilation cross section, from cross correlation
  with 2MRS and for both the boost models~\cite{Gao2012,
  Sanchez-Conde2014}. The dashed curves are the median of the
  distribution, while the solid curves are the same as shown in
  Fig.~\ref{fig:sigmav_2MRS}.}
  \label{fig:sigmav_2MRS_simulated}
 \end{center}
\end{figure}

The results in the case of cross correlation with 2MRS are shown in
Fig.~\ref{fig:sigmav_2MRS_simulated} as thick and thin bands, with which
we show 68\% and 95\% containment intervals of the 95\% credible
sensitivities.
The median of the distribution is shown as the dashed curves, while
the solid curves are the same as the ones shown in
Fig.~\ref{fig:sigmav_2MRS}.
We find that fluctuation of cross-correlation data yields difference of
the upper limits on the annihilation cross section by almost one order
of magnitude.

\section{Conclusions}
\label{sec:conclusions}

As was pointed out in Ref.~\cite{ABK}, taking cross correlation between
the gamma-ray background (due to Fermi-LAT) and the local galaxy
distribution such as 2MASS catalogs provides very efficient way of
constraining dark matter annihilation cross section.
This is because the dark matter annihilation contributes more to the
gamma-ray background from the local Universe at lower redshifts.

In this paper, we aimed at making further theoretical progress.
Most importantly, since galaxies in major catalogs are assigned with
either spectroscopic or photometric redshifts, one could further use
this information to disentangle dark matter from other astrophysical
sources.
Taking such a tomographic approach, we divided the 2MASS catalogs into
more than one (up to ten) redshift bins.
We found that we could already improve the sensitivities to the
annihilation cross section by a factor of a few to several, if we
divided the catalogs into two or three redshift slices.
Beyond four slices, the improvement is possible but modest.
If the dark matter halos contain a large number of substructure yielding
a large boost of the annihilation signals (e.g., \cite{Gao2012}), then
the canonical annihilation cross section would be probed for the dark
matter less massive than $\sim$700~GeV.
For a more modest scenario with less substructure boost (e.g.,
\cite{Sanchez-Conde2014}), the sensitivity reaches around
$\sim$10$^{-25}$~cm$^{3}$~s$^{-1}$ for dark matter with masses of tens
to a few hundreds of GeV.

For these estimates, we developed theory and included the shot-noise
term in the cross-correlation power spectrum, which comes from the fact
that descrete point sources in the galaxy catalogs contribute partly to
the gamma-ray background.
The cross-power spectrum between density squared and galaxy distribution
as well as the galaxy power spectrum were computed with the updated halo
model adopting the halo occupation distribution (see also
Ref.~\cite{Fornengo2014}), and we found considerable differences in
relevant angular scales.

We based our sensitivity estimate on the Bayesian statistics and
Markov-Chain Monte Carlo simulations.
This enables us to adopt priors of theoretical parameters, excluding
unphysical values (such as negative annihilation cross section) from the
beginning.
This turned out to be a major improvement compared with the simplistic
estimates based on the Fisher matrix adopted in Ref.~\cite{ABK} as shown
in Fig.~\ref{fig:sigmav_2MRS}.
We also found that the statistical fluctuation of data would yield
uncertainty band of one order of magnitude for the sensitivity estimates
(Fig.~\ref{fig:sigmav_2MRS_simulated}).

\section*{Acknowledgments}

This work was supported by the Netherlands Organization for Scientific
Research (NWO) through a Vidi grant.

\appendix

\section{Shot noise in the cross-correlation power spectrum}
\label{app:shot noise}

If the gamma-ray sources are point-like and if they are correlated with
catalog galaxies, there is a shot noise in both the auto- and
cross-correlation power spectra, although that for the latter was not
taken into account in Ref.~\cite{ABK}.
We first write Eq.~(\ref{eq:I_X}) by explicitly showing descrete nature
of astrophysical point sources $X$ as
\begin{equation}
 I_X(\hat{\bm n}) = \int d\chi W_X(z)
  \frac{1}{\langle n_X (z) \rangle} \sum_{i \in X} \delta_{\rm D}^3 (\bm
  x - \bm x_i),
  \label{eq:I_X delta}
\end{equation}
where $\bm x$ represents the comoving coordinates, and $\delta_{\rm
D}^N$ is the $N$-dimensional Dirac delta function.
Similarly, the surface density of galaxies (that are represented by g)
is
\begin{equation}
 \Sigma_{\rm g}(\hat{\bm n}) = \frac{4\pi f_{\rm g}}{N_{\rm g}}
  \int d\chi \chi^2 \sum_{j \in {\rm g}} \delta_{\rm D}^3 (\bm x - \bm
  x_j),
\end{equation}
where $N_{\rm g}$ is the total number of galaxies found in sky fraction
$f_{\rm g}$.
This surface density is normalized to give one after taking ensemble
average, $\langle \Sigma_{\rm g} \rangle = 1$.

Two-point angular cross correlation between $\delta I_X = I_X - \langle
I_X \rangle$ and $\delta \Sigma_{\rm g} = \Sigma_{\rm g} - \langle
\Sigma_{\rm g} \rangle$ is then
\begin{eqnarray}
 C_{X, {\rm g}}(\theta) &=& \langle \delta I_X(\hat {\bm n}_1) \delta
  \Sigma_{\rm g}(\hat{\bm n}_2)\rangle
  \nonumber\\
 &=& \frac{4\pi f_{\rm g}}{N_{\rm g}}\int d\chi_1
  \frac{W_X(z_1)}{\langle n_X(z_1)\rangle}
  \nonumber\\&&{}\times
  \int d\chi_2 \chi_2^2 \left[
  \langle n_X(\bm x_1) n_{\rm g}(\bm x_2) \rangle
  - \langle n_X(z_1)\rangle \langle n_{\rm g}(z_2) \rangle\right]
  ,
\end{eqnarray}
where $\cos \theta = \hat{\bm n}_1\cdot \hat{\bm n}_2$.
To evaluate the ensemble average of the product of the densities (i.e.,
delta functions), we assume that one of the two source classes ($X$ or
g) is completely included in the other; i.e., $X \subset {\rm g}$ or $X
\supset {\rm g}$.
For example, in the case of SFGs, it is natural to expect that they form
sub-sample
of the local 2MASS galaxies at low redshifts (where 2MASS survey is
complete), while they include 2MASS galaxies at high redshifts.
In the case that $X \subset {\rm g}$, then we evaluate it as
\begin{eqnarray}
 \langle n_X(\bm x_1) n_{\rm g}(\bm x_2)\rangle
 &=& \left\langle \sum_{i \in X} \delta_{\rm D}^3 (\bm x_1 - \bm x_i)
      \sum_{j = i} \delta_{\rm D}^3 (\bm x_2 - \bm x_j) \right\rangle
 \nonumber\\&&{}
 +  \left\langle \sum_{i \in X} \delta_{\rm D}^3 (\bm x_1 - \bm x_i)
     \sum_{j \ne i} \delta_{\rm D}^3 (\bm x_2 - \bm x_j) \right\rangle
 \nonumber\\ &=&
  \langle n_X (z_1) \rangle \delta_{\rm D}^3 (\bm x_1 - \bm x_2)
  + \langle n_X(z_1) \rangle \langle n_{\rm g}(z_2) \rangle [1 + \xi_{X,
  {\rm g}}(\bm x_1 - \bm x_2)],
  \label{eq:ensemble average of density correlation}
\end{eqnarray}
where $\xi_{X, {\rm g}}$ is the two-point correlation function between
$X$ and the galaxies.
In order to include the case that $X \supset {\rm g}$, we generalize the
formula by replacing $\langle n_X(z_1) \rangle$ with $\min[\langle
n_{X}(z_1) \rangle, \langle n_{\rm g}(z_1) \rangle]$ in the first
term.\footnote{In the case that there is partial overlap between $X$ and
g, this factor should be the density of sources in the overlap.}

One can convert the correlation function $C_{X, {\rm g}}(\theta)$ to the
angular power spectrum $C_\ell^{X, {\rm g}}$ through two-dimensional
Fourier transform (flat-sky approximation valid for small angular
scales), and the procedure is the same as the one summarized in
Appendix~A of Ref.~\cite{Ando2007}.
As the result, we obtain
\begin{equation}
 C_\ell^{X, {\rm g}} = \int \frac{d\chi}{\chi^2}W_X(z) W_{\rm g}(z)
  \left\{\frac{1}{\max[\langle n_X(z)\rangle, \langle n_{\rm g}(z)
   \rangle]} + P_{X, {\rm g}}\left(\frac{\ell}{\chi},
			      z\right)\right\},
\end{equation}
where $W_{\rm g}(z) = 4\pi f_{\rm sky, g} \chi^2 \langle n_{\rm g}(z)
\rangle / N_{\rm g}$ and $P_{X, {\rm g}}(k, z)$ is the power spectrum
cross-correlated between distributions of $X$ and the galaxies (i.e.,
Fourier transform of $\xi_{X, {\rm g}}$).
The first term represents the shot noise coming from the descrete nature
of the point sources, while the second is the correlation term that is
visited in the following section.

\section{Halo occupation distribution and galaxy power spectrum}
\label{sec:App B}

\subsection{Galaxy power spectrum}
\label{sub:App B1}

We now derive the cross-correlation power spectrum between the point
source class $X$ and the galaxies, $P_{X, {\rm g}}(k,z)$.
As for $X$, we consider both the SFGs and blazars.
For SFGs, the quantity is simply the galaxy power spectrum, $P_{\rm
g}(k, z)$.
One can assume that this is proportional to the matter power spectrum
with a constant bias as adopted in Ref.~\cite{ABK}.
Although such an approach is reasonable for large scales where the
density fluctuation is in the linear regime, it will significantly
differ from the galaxy power spectrum at small scales as physics of
galaxy formation starts playing an important role.

One could still phenomenologically estimate the galaxy power spectrum in
the context of halo model~\cite{Seljak2000, Cooray2002}.
The halo occupation distribution (HOD) is a probability distribution function
of having $N$ galaxies in a host halo with mass $M$, $P_N(M)$ (see,
e.g., Ref.~\cite{Seljak2000, Zheng2005} and references therein).
If this quantity and also the distribution of galaxies in a host halo
are known, then one is able to compute the galaxy power spectrum.
Based on numerical simulations as well as semi-analytic models of galaxy
formation, Ref.~\cite{Zheng2005} found that the mean galaxy number in
the host halo with mass $M$, $\langle N_{\rm g} | M \rangle$, is well
fitted by
\begin{eqnarray}
 \langle N_{\rm g} | M \rangle &=& \langle N_{\rm cen} | M \rangle +
  \langle N_{\rm sat} | M \rangle ,
  \label{eq:HOD}\\
 \langle N_{\rm cen} | M \rangle &=& \frac{1}{2}
  \left[1 + {\rm erf}\left(\frac{\log M - \log M_{\rm min}}{\sigma_{\log
		      M}}\right)\right] ,
  \label{eq:HOD_cen}\\
 \langle N_{\rm sat} | M \rangle &=& \left(\frac{M -
				      M_0}{M_1}\right)^\alpha ,
 \label{eq:HOD_sat}
\end{eqnarray}
where `cen' and `sat' represent the central and satellite galaxies,
respectively, erf is the error function, and the parameters are $\log
(M_{\rm min} / M_\odot) = 11.68$, $\sigma_{\log M} = 0.15$, $\log
(M_0/M_\odot) = 11.86$, $\log (M_1 / M_\odot) = 13.0$, and $\alpha =
1.02$.
Note that the number of the central galaxy saturates at 1 for high-mass
halos, while that of the satellite galaxies grow almost linearly with the
halo mass.
The galaxy density is obtained by integrating the number of galaxies
weighed by the mass function,
\begin{equation}
 \langle n_{\rm g}(z) \rangle = \int dM \frac{dn(M,z)}{dM}
  \langle N_{\rm g} | M \rangle.
  \label{eq:n_g}
\end{equation}

\begin{figure}[t]
 \begin{center}
  \includegraphics[width=8.5cm]{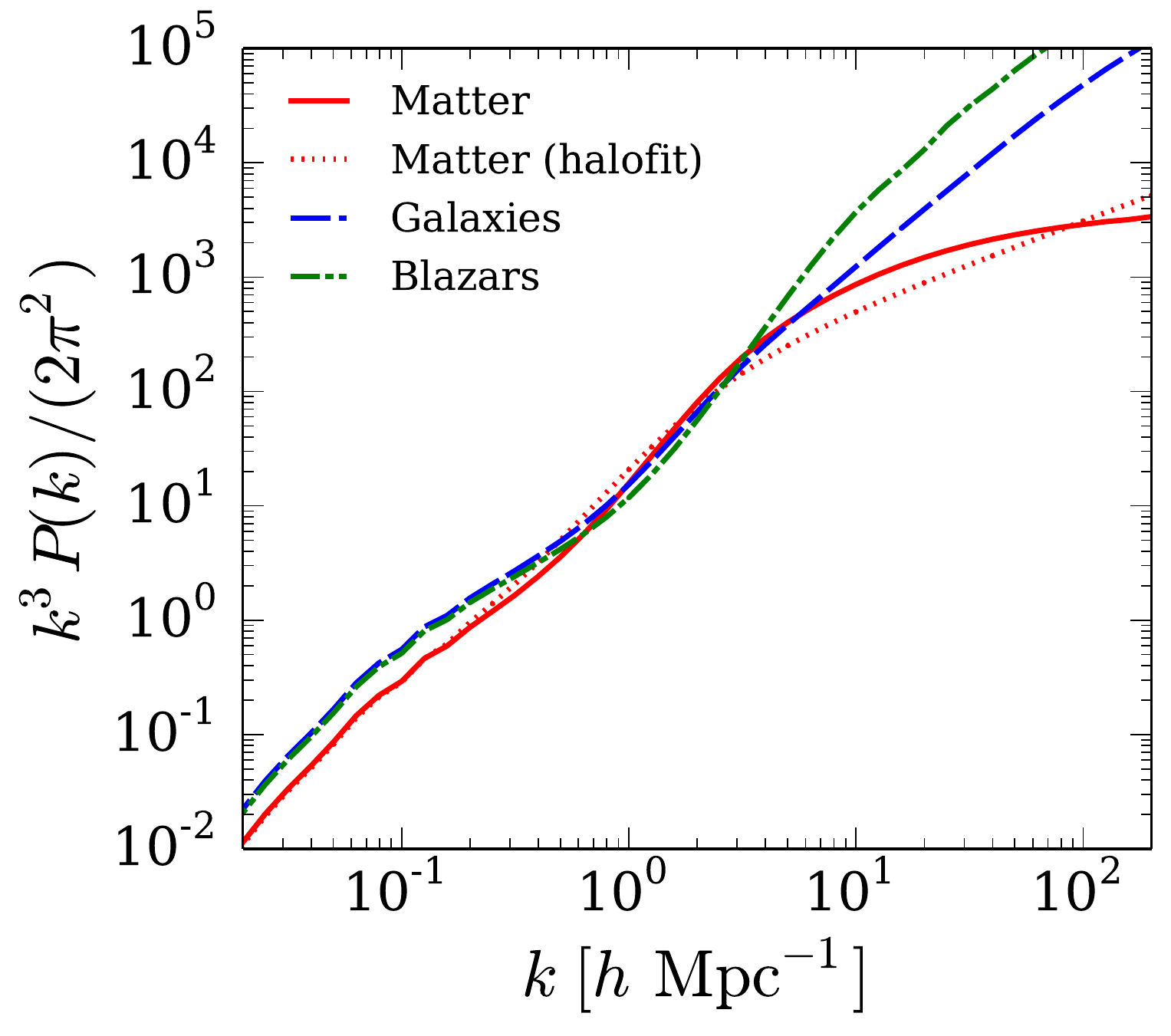}
  \caption{Power spectra of matter (solid), galaxies (dashed), and
  blazars (dot-dashed) due to halo model at $z = 0$. The matter power
  spectrum due to `halofit'~\cite{Takahashi2012} is also shown for
  comparison (dotted).}
  \label{fig:powerspectrum}
 \end{center}
\end{figure}

We implement the halo model computation by treating central and
satellite galaxies separately.
The galaxy power spectrum as the result is divided into one-halo and
two-halo terms as follows (e.g., \cite{Seljak2000, Sheth2003}):
\begin{eqnarray}
 P_{\rm g}(k, z) &=& P_{\rm g}^{\rm 1h}(k, z) + P_{\rm g}^{\rm 2h}(k,
  z),
  \label{eq:P_g}\\
 P_{\rm g}^{\rm 1h}(k, z) &=& \int dM \frac{dn(M, z)}{dM}
  \frac{1}{\langle n_{\rm g}(z) \rangle^2}
  \left[2 \langle N_{\rm cen} N_{\rm sat} | M \rangle \tilde
   u_{\rm sat}(k | M) 
   \right.\nonumber\\&&{}\left. + 
   \langle N_{\rm sat}(N_{\rm sat} - 1) | M \rangle |\tilde u_{\rm
   sat}(k|M)|^2\right],
   \label{eq:P_g 1h}\\
 P_{\rm g}^{\rm 2h}(k, z) &=& b_{\rm 1, g}^2(z) P_{\rm lin}(k, z),
  \label{eq:P_g 2h}
\end{eqnarray}
where $\tilde u_{\rm sat}(k | M)$ is the Fourier transform of the
normalized satellite distribution in a halo with the mass $M$, $u_{\rm
sat}(r|M)$, for which we assume the NFW profile, and $b_{1, {\rm g}}$ is
the linear bias, for which we adopt 1.4 independent of
redshifts~\cite{Davis2011}.
Here we assumed that the central galaxies locate at the center of the host
halo, and therefore, the auto-correlation of the central galaxy simply
yields a shot noise that is not included here but discussed in
Appendix~\ref{app:shot noise}.
The first term of Eq.~(\ref{eq:P_g 1h}) represents the correlation
between the central and one of the satellite galaxies, while the second
term represents that between two of the satellites.
Remembering that $N_{\rm sat} > 0$ only if $N_{\rm cen} = 1$, we have
$\langle N_{\rm cen} N_{\rm sat} | M \rangle = \langle N_{\rm sat} | M
\rangle$.
Also the HOD of satellite galaxies is well approximated to the Poisson
distribution~\cite{Zheng2005}, and thus we have $\langle N_{\rm sat}
(N_{\rm sat} - 1)| M \rangle = \langle N_{\rm sat} | M \rangle^2$.
With these and Eqs.~(\ref{eq:HOD})--(\ref{eq:HOD_sat}), we compute the
galaxy power spectrum through Eqs.~(\ref{eq:P_g})--(\ref{eq:P_g 2h}).

Figure~\ref{fig:powerspectrum} shows the galaxy power spectrum $P_{\rm
g} (k)$ at $z = 0$, and the comparison with the matter power spectrum
using the same halo model (e.g., \cite{Seljak2000, Cooray2002}) and a
`halofit' model~\cite{Smith2003, Takahashi2012}.
The linear power spectrum as well as the halofit model were computed
with the {\tt CAMB} numerical code.\footnote{\url{http://camb.info}}
We find substantial deviation of the galaxy power spectrum from that
for matter especially at scales smaller than $\sim$0.1~$h^{-1}\, {\rm
Mpc}$.

\subsection{Blazar-galaxy cross-power spectrum}
\label{sub:App B2}

The HOD for the gamma-ray blazar population is
hardly known observationally.
It is, however, possible to use the knowledge from other wavebands such
as X rays in order to study it.
Although clustering properties of X-ray AGNs are certainly less studied
compared with those of galaxies, it is found that they selectively
live in $\sim$10$^{13.1} h^{-1} M_\odot$ dark matter
halos~\cite{Allevato2011, Cappelluti2012}.
We here assume that such a property is common for the AGNs that are also
bright in gamma rays, and also that such AGNs only reside at the center
of the host dark matter halos.
With these approximations, we write the blazar-galaxy cross-power
spectrum $P_{\rm b, g}(k, z)$ as
\begin{eqnarray}
 P_{\rm b,g}(k,z) &=& P_{\rm b,g}^{\rm 1h}(k,z) + P_{\rm b,g}^{\rm
  2h}(k,z),\\
 P_{\rm b,g}^{\rm 1h}(k,z) &=&
  \frac{\langle N_{\rm sat} | M_{\rm b} \rangle}{\langle n_{\rm g}(z)\rangle}
  \tilde u_{\rm sat}(k | M_{\rm b}),\\
 P_{\rm b,g}^{\rm 2h}(k,z) &=& b_1(M_{\rm b}, z) b_{\rm 1,g}(z) P_{\rm
  lin}(k,z),
\end{eqnarray}
where $M_{\rm b} = 10^{13.1} h^{-1} M_\odot$ is the mass of the dark
matter halos that host blazars.
The one-halo term takes correlation between a central AGN and satellite
galaxies, while the correlation with a central galaxy yields shot noise
that was treated in Appendix~\ref{app:shot noise}.
The former corresponds to the first term of Eq.~(\ref{eq:P_g 1h}), but
divided by 2 and computed with a very sharp mass function, $dn / dM
\propto \delta_{\rm D}(M - M_{\rm b})$.

We show the blazar power spectrum in Fig.~\ref{fig:powerspectrum}, and
find that they cluster even more strongly than galaxies at small scales.

\section{Cross correlation between density squared and galaxy
 distribution}
\label{sec:App C}

In the halo model, all the dark matter particles are confined in halos.
The central position and mass of the halo are represented by $\bm x_i$
and $M_i$, respectively, and the matter density squared is then written
as
\begin{equation}
 \rho_{\rm m}^2 (\bm x) = \int d^3 x^\prime \int dM \sum_i \delta_{\rm
  D}^3 (\bm x^\prime - \bm x_i) \delta_{\rm D}(M - M_i) \mathcal J(M,
  z^\prime) u_{\delta^2}(\bm x - \bm x^\prime | M),
\end{equation}
where $u_{\delta^2}(\bm x | M)$ is the profile of density squared of a
halo with mass $M$, which is normalized to unity after volume
integration.
Recalling that the ensemble average of the product of delta functions
gives the halo mass function:
\begin{equation}
 \left\langle \sum_i \delta_{\rm D}^3(\bm x - \bm x_i)
  \delta_{\rm D}(M - M_i)\right\rangle = \frac{dn(M, z)}{dM},
\end{equation}
we recover Eq.~(\ref{eq:variance}).

For the number density of the galaxies, instead of using the sum of the
delta functions of the coordinates [Eq.~(\ref{eq:I_X delta})], we regard
galaxies as particles smoothly distributed around the halo at $\bm x_j$,
following $u_{\rm g}$:
\begin{equation}
 n_{\rm g}(\bm x) = \int d^3x^\prime \int dM \sum_j \delta_{\rm D}^3
  (\bm x^\prime - \bm x_j) \delta_{\rm D}(M - M_j) N_{\rm g}(M) u_{\rm
  g}(\bm x - \bm x^\prime | M).
\end{equation}
Again, by taking the ensemble average, one recovers Eq.~(\ref{eq:n_g}).

\begin{figure}[t]
 \begin{center}
  \includegraphics[width=8.5cm]{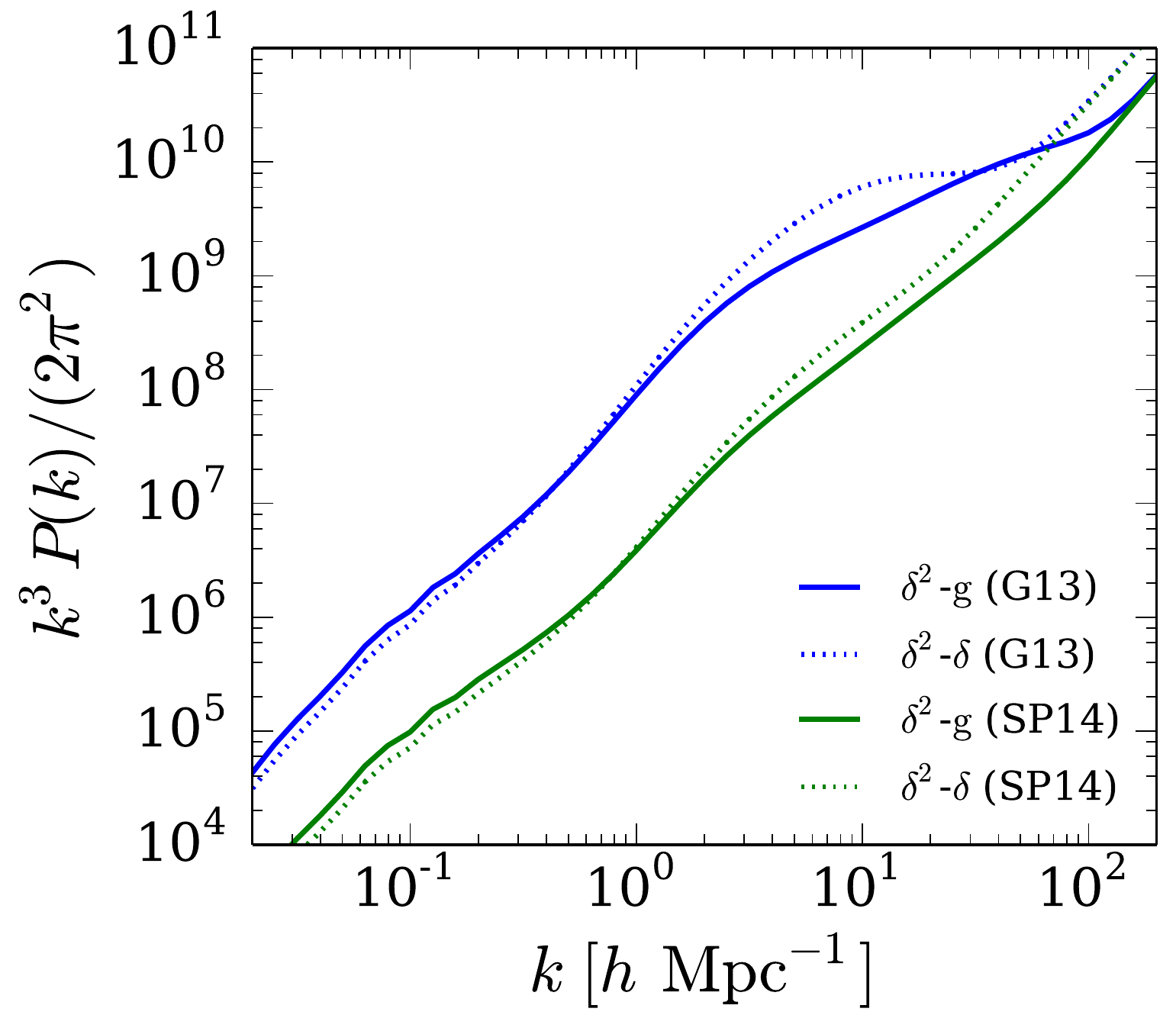}
  \caption{Cross-power spectrum between density squared and galaxy
  distribution (solid) or density (dotted) at $z = 0$. Results for two
  boost models are shown; \cite{Gao2012} for the upper curves (labeled
  as G13) and \cite{Sanchez-Conde2014} for the lower (SP14).}
  \label{fig:cross_powerspectrum}
 \end{center}
\end{figure}

Following the same argument as in Eq.~(\ref{eq:ensemble average of
density correlation}), the two-point cross-correlation function between
$\delta \rho_{\rm m}^2 = \rho_{\rm m}^2 - \langle \rho_{\rm m}^2
\rangle$ and $\delta n_{\rm g}$ is written as
\begin{eqnarray}
 \langle \delta \rho_{\rm m}^2(\bm x_1) \delta n_{\rm g}(\bm x_2)\rangle
  &\equiv&
  \langle \rho_{\rm m} \rangle^2 \langle n_{\rm
  g}(z_2)\rangle \xi_{\delta^2, {\rm g}}(\bm x_1 - \bm x_2) \nonumber\\
  &=& \int d^3x_1^\prime \int dM_1 \frac{dn(M_1, z_1)}{dM_1} \mathcal
  J(M_1, z_1) u_{\delta^2}(\bm
  x_1 - \bm x_1^\prime | M_1)
  \nonumber\\&&{}\times
  \langle N_{\rm  g} | M_1 \rangle u_{\rm
  g}(\bm x_2 - \bm x_1^\prime | M_1)
  \nonumber\\&&{}
  + \int d^3 x_1^\prime \int dM_1 \int d^3 x_2^\prime \int dM_2
  \frac{dn(M_1, z_1)}{dM_1}\frac{dn(M_2, z_2)}{dM_2}
  \nonumber\\&&{}\times
  \mathcal J(M_1, z_1) u_{\delta^2}(\bm x_1 - \bm x_1^\prime | M_1)
  \langle N_{\rm g}|M_2 \rangle u_{\rm g}(\bm x_2 - \bm x_2^\prime |
  M_2)
  \nonumber\\&&{}\times
  \xi_{\rm hh}(\bm x_1^\prime - \bm x_2^\prime | M_1, M_2),
  \label{eq:xi_delta2_gal}
\end{eqnarray}
where $\xi_{\rm hh}$ is the two-point correlation function of the halo
seeds.
In the following, we approximate it as $\xi_{\rm hh}(r|M_1,M_2) \approx
b_1(M_1) b_2(M_2) \xi_{\rm lin}(r)$, where $\xi_{\rm lin}$ is the linear
correlation function.
The cross-power spectrum, defined as the Fourier transform of
$\xi_{\delta^2, {\rm g}}(r)$ [Eq.~(\ref{eq:xi_delta2_gal})], is
derived, after straightforward algebra, as
\begin{eqnarray}
 P_{\delta^2, {\rm g}}(k, z) &=& P_{\delta^2, {\rm g}}^{\rm 1h}(k, z) +
  P_{\delta^2, {\rm g}}^{\rm 2h}(k, z), \\
 P_{\delta^2, {\rm g}}^{\rm 1h}(k, z) &=&
  \int dM \frac{dn(M,z)}{dM}
  \frac{\mathcal J(M, z)}{(\Omega_{\rm m} \rho_{\rm c})^2}
  \frac{\langle N_{\rm g}| M \rangle}{\langle n_{\rm g}(z)\rangle}
  \tilde u_{\delta^2}(k|M) \tilde u_{\rm g}(k|M),\\
 P_{\delta^2, {\rm g}}^{\rm 2h}(k, z) &=&
  \left[\int dM \frac{dn(M,z)}{dM}\frac{\mathcal J(M, z)}{(\Omega_{\rm
  m}\rho_{\rm c})^2} \tilde u_{\delta^2}(k|M)
  b_1(M, z)\right]
  \nonumber\\&&{}\times
  \left[\int dM^\prime \frac{dn(M^\prime, z)}{dM^\prime}
  \frac{\langle N_{\rm g}| M^\prime \rangle}{\langle n_{\rm g}(z)\rangle}
   \tilde u_{\rm g}(k|M^\prime) b_1(M^\prime, z)\right] P_{\rm lin}(k, z),
\end{eqnarray}
where $\tilde u_{\delta^2}(k)$ is the Fourier transform of
$u_{\delta^2}(r)$, and
\begin{equation}
 \langle N_{\rm g} | M \rangle \tilde u_{\rm g}(k | M) = \langle
  N_{\rm cen} | M \rangle + \langle N_{\rm sat} | M \rangle \tilde
  u_{\rm sat}(k|M).
\end{equation}
In Fig.~\ref{fig:cross_powerspectrum}, we show $P_{\delta^2, {\rm g}}(k,
0)$ for both the boost models~\cite{Gao2012, Sanchez-Conde2014}.
We also compare the results with $P_{\delta^2, \delta}(k, 0)$ within the
same framework of the halo model, which was adopted in Ref.~\cite{ABK}.
At small scales, $\delta^2$-g correlation is smaller even though it is
very similar to $\delta^2$-$\delta$ power spectrum.

Finally, after projection for which the same procedure is used as in
Appendix~\ref{app:shot noise}, the angular cross-power spectrum between
dark matter annihilation and the galaxy distribution is obtained with
\begin{equation}
 C_\ell^{\rm dm, g} = \int\frac{d\chi}{\chi^2}W_{\rm dm}(z) W_{\rm g}(z)
  P_{\delta^2, {\rm g}}\left(\frac{\ell}{\chi}, z\right).
\end{equation}

\bibliographystyle{JHEP}
\bibliography{refs.bib}

\end{document}